\documentclass[useAMS,usenatbib]{mn2e}

\usepackage{psfig}
\usepackage{graphicx}
\usepackage{times}
\usepackage{natbib}

\newif\ifAMStwofonts
\AMStwofontstrue

\newcommand{\be}{\begin{equation}}
\newcommand{\ee}{\end{equation}}
\newcommand{\ba}{\begin{eqnarray}}
\newcommand{\ea}{\end{eqnarray}}
\newcommand{\brr}{\begin{array}}
\newcommand{\err}{\end{array}}
\newcommand{\bc}{\begin{center}}
\newcommand{\ec}{\end{center}}
\newcommand{\hm}{\,h^{-1}{\rm Mpc}}
\newcommand{\hk}{\,h^{-1}{\rm kpc}}
\newcommand{\msun}{\,h^{-1}{\rm M}_\odot}

\newcommand{\rvir}{\mbox{$R_{\rmn{vir}}$}}
\newcommand{\mvir}{\mbox{$M_{\rmn{vir}}$}}
\newcommand{\tmw}{\mbox{$T_{\rmn{mw}}$}}

\newcommand{\vel}{\,{\rm km\,s^{-1}}}

\newcommand{\mincir}{\raise
  -2.truept\hbox{\rlap{\hbox{$\sim$}}\raise5.truept \hbox{$<$}\ }}
\newcommand{\magcir}{\raise
  -2.truept\hbox{\rlap{\hbox{$\sim$}}\raise5.truept \hbox{$>$}\ }}
\newcommand{\siml}{\raise
  -2.truept\hbox{\rlap{\hbox{$\sim$}}\raise5.truept \hbox{$<$}\ }}
\newcommand{\simg}{\raise
  -2.truept\hbox{\rlap{\hbox{$\sim$}}\raise5.truept \hbox{$>$}\ }}

\newcommand{\tarkin}{Set~\#1}
\newcommand{\hutt}{Set~\#2}

\title[Hot and Cooled baryons in SPH simulations
of galaxy clusters] {Hot and Cooled baryons in SPH simulations
of galaxy clusters: physics and numerics}

\author[S. Borgani, et al.]
{S. Borgani$^{1,2}$, K. Dolag$^3$, G. Murante$^{4,1}$, L.-M. Cheng$^5$,
  V. Springel$^3$, \\ ~\\
\LARGE{\rm A. Diaferio$^6$, L. Moscardini$^7$, G. Tormen$^8$,
  L. Tornatore$^{9,2}$ \& P. Tozzi$^{10,2}$}\\ ~\\
$^1$ Dipartimento di Astronomia dell'Universit\`a di Trieste, via
  Tiepolo 11, I-34131 Trieste, Italy (borgani@ts.astro.it)\\
$^2$ INFN -- National Institute for Nuclear Physics, Trieste,
  Italy\\ 
$^3$ Max-Planck-Institut f\"ur Astrophysik, Karl-Schwarzschild Strasse
  1, Garching bei M\"unchen, Germany (kdolag,volker@mpa-garching.mpg.de)\\
$^4$ INAF, Osservatorio Astronomico di Torino, Strada Osservatorio 20,
  I-10025 Pino Torinese, Italy (giuseppe@to.astro.it)\\
$^5$ Institute of Theoretical Physics, Chinese Academy of Sciences,
Zhong Guan Cun South 4th Street, 1A P.O.Box 2735, Beijing 100080,
China (clm@itp.ac.cn)\\
$^6$ Dipartimento di Fisica Generale ``Amedeo Avogadro'', Universit\'a
  degli Studi di Torino, Torino, Italy (diaferio@ph.unito.it) \\
$^7$ Dipartimento di Astronomia, Universit\`a di Bologna, via Ranzani
  1, I-40127 Bologna, Italy (lauro.moscardini@unibo.it)\\
$^8$ Dipartimento di Astronomia, Universit\`a di Padova, vicolo
  dell'Osservatorio 2, I-35122 Padova, Italy (tormen@pd.astro.it)\\
$^9$ SISSA/International School for Advanced Studies, via Beirut 4,
34100 Trieste, Italy (torna@sissa.it)\\
$^{10}$ INAF, Osservatorio Astronomico di Trieste, via Tiepolo 11,
  I-34131 Trieste, Italy (tozzi@ts.astro.it)}
\begin{document}

\date{Accepted ???. Received ???; in original form ???}

%\pagerange{\pageref{firstpage}--\pageref{lastpage}} \pubyear{0000}

\maketitle

\label{firstpage}

\begin{abstract}
  We discuss an extended set of Tree+SPH simulations of the formation of
  clusters of galaxies, with the goal of investigating the interplay between
  numerical resolution effects and star--formation/feedback processes.  Our
  simulations were all carried out in a concordance $\Lambda$CDM cosmology and
  include radiative cooling, star formation, and energy feedback from galactic
  winds.  The simulated clusters span the mass range $\mvir \simeq
  (0.1-2.3)\times 10^{15}\msun$, with mass resolution varying by several
  decades. At the highest achieved resolution, the mass of gas particles is
  $m_{\rm gas}\simeq 1.5\times 10^7\msun$, which allows us to resolve the
  virial region of a Virgo--like cluster with more than 2 million gas
  particles and with at least as many dark--matter (DM) particles.  Our
  resolution study confirms that, in the absence of an efficient feedback
  mechanism, runaway cooling leads to about 35 per cent of baryons in clusters
  to be locked up in long lived stars at our highest resolution, with no
  evidence of convergence.  However, including feedback causes the fraction of
  cooled baryons to converge at about 15 per cent already at modest
  resolution, which is much closer to the typical values inferred from
  observational data.  Feedback also stabilizes other gas--related quantities,
  such as radial profiles of entropy, gas density and temperature, against
  variations due to changes in resolution.
  Besides effects of mass resolution, we also investigate the influence of the
  gravitational force softening length, and that of numerical heating of the
  gas induced by two-body encounters between DM and lighter gas
  particles. We also show
  that simulations where more DM than gas particles are used, such that
  $m_{\rm gas}\simeq m_{\rm DM}$, show a significantly enhanced efficiency of
  star formation at $z\magcir 3$, but they accurately reproduce at $z=0$ the
  fraction of cooled gas and the thermodynamical properties of the
  intra--cluster gas.  Our results are important for establishing and
  delineating the regime of numerical reliability of the present generation of
  hydrodynamical simulations of galaxy clusters.
\end{abstract}


\section{Introduction} \label{sec:intro}

Within the hierarchy of cosmic structures, clusters of galaxies mark
an interesting transition between the large--scale regime, where the
dynamics is dominated by gravity, and the small--scale regime, where
complex astrophysical processes, such as gas cooling, star formation
and energetic feedback processes, control the formation and evolution
of galaxies \citep[e.g.][ for recent
reviews]{2002ARA&A..40..539R,2005RvMP...77..207V}. If gravity were the
only player at the scales relevant for galaxy clusters, then the
overall thermal content of baryons would be completely determined by
the processes of adiabatic compression and shock heating within the
dark--matter (DM) dominated potential wells
\citep[e.g.][]{1986MNRAS.222..323K,2001ApJ...546...63T}. Since gravity
has no characteristic scales, this scenario predicts that galaxy
clusters and groups of different mass should appear as scaled versions
of each other, with only weak residual trends due to a variation of
the halo concentration with mass. However, a number of observational
facts highlight that the evolution of cosmic baryons within clusters
must be influenced by physical processes other than gravity.  First,
scaling relations between X--ray observables demonstrate that groups
and poor clusters have a lower content of diffuse hot gas than rich
clusters, and that they lie on a comparatively higher adiabat
\citep[e.g.][]{1998ApJ...504...27M,1999MNRAS.305..631A,2002A&A...391..841E,2003MNRAS.343..331P,2004MNRAS.350.1511O}.
Second, the bulk of the gas in central cluster regions lies at a
temperature which is never observed to fall below 1/3--1/2 of the
virial temperature, despite the fact that the cooling time of this gas
is much shorter than the Hubble time
\citep[e.g.][]{2001A&A...365L.104P,2001ApJ...560..194M,2002A&A...382..804B}.
Third, the fraction of baryons in the stellar phase within clusters is
consistently rather small, around $\sim 10$ per cent
\citep[e.g.][]{2001MNRAS.326.1228B}, with a tendency for poor clusters
and groups to have a somewhat higher amount of stars
\citep[e.g.][]{2003ApJ...591..749L,2004AJ....128.1078R}.

These observational evidences indicate the existence of a delicate interplay
between the physical processes which determine the evolution of the
inter--galactic (IGM) and intra--cluster (ICM) media. Including hydrodynamical
effects beyond ordinary gas dynamics, such as gas cooling, star formation and
associated feedback processes, in simulation models is however met with
substantial difficulty.  While analytical and semi--analytical calculations
can provide very useful guidelines for the expected effects
\citep[e.g.][]{2001ApJ...546...63T,2001MNRAS.325..497B,2002MNRAS.330..329B,2003ApJ...593..272V,2005ApJ...619...60L},
we nevertheless rely on such simulation models, because the non-linearity of
the physics and the geometrical complexity of cluster dynamics can usually
only be captured in full by direct simulation.  This makes it extremely
important to obtain an in-depth understanding of the robustness of the
numerical methods employed, of the assumptions behind them, and of the
sensitivity of numerical predictions for observable quantities on the adopted
numerical parameters and approximations.

One important validation test is to compare different simulation codes
and different schemes for solving hydrodynamics with each other. For a
single code, additional control tests that scrutinize the robustness
of the predictions against resolution and different implementations of
sub--resolution processes are highly desirable as well. An example for
the importance of such tests is represented by radiative cooling,
which is well known to exhibit a runaway character: cooling leads to
an increase of the gas density, which, in turn, increases the cooling
efficiency. Consistently, hydrodynamical cosmological simulations have
shown that the fraction of gas that cools down and becomes available
for star formation is much larger than indicated by observations
\citep[e.g.][]{1993ApJ...412..455K,1998ApJ...507...16S,2000MNRAS.317.1029P,2000ApJ...536..623L,2002MNRAS.335..762Y},
a tendency that increases with resolution
\citep[e.g.][]{2001MNRAS.326.1228B,2003MNRAS.342.1025T}. Solving this
problem apparently requires the introduction of suitable feedback
physics which reduces the fraction of cooled gas and stabilizes its
value against numerical resolution.

So far, detailed comparisons between different cosmological hydrodynamical
codes have been restricted to the case of non--radiative simulations
\citep[e.g.][]{1994ApJ...430...83K,1999ApJ...525..554F,2005ApJS..160....1O}.
While these comparisons have shown a reasonable level of agreement, some
sizeable differences have been found in the profiles of thermodynamic
quantities (e.g., entropy) when simulations of galaxy clusters performed with
Eulerian and Lagrangian codes were compared \citep{1999ApJ...525..554F}.  A
limited number of studies have been presented so far which were aimed at
discussing the effect of resolution and different implementations of
cooling/star--formation physics within a single code. For instance,
\cite{2002MNRAS.330..113K} compared different implementations of star
formation. They concluded that different prescriptions provide broadly
consistent results, although the amount of cooled gas is always much higher
than observed. They also verified that this overcooling can be partly ameliorated
by resorting to either kinetic or thermal feedback.

Besides checking the influence of different parameterizations of physical
effects, it is crucial to have purely numerical effects well under control,
such as those originating from mass resolution and force softening.  Detailed
studies have been presented about the determination of an optimum softening
for purely collisionless simulations
\citep[e.g.][]{2003MNRAS.338...14P,2005astro.ph..7237Z}, usually
based on a compromise between the desire of obtaining an accurate estimate of
the accelerations together with high spatial resolution, and at the same time, the need
to suppress two-body relaxation and the formation of bound particle pairs if
too small a softening is chosen \citep[e.g.][]{1992MNRAS.257...11T}.

The situation is more complicated for hydrodynamical simulations, where energy
can be spuriously transferred from the collisionless to the collisional
component, thereby affecting the evolution of the gas.  As
\cite{1997MNRAS.288..545S} have shown, this can happen when the gas particles
are substantially lighter than the dark matter particles, such that the former
receive a systematic energy transfer in two--body encounters, leading to
artificial heating of the gas.  As a result, a lower--limit exists for the
required mass resolution in order to provide a correct description of gas
cooling within halos of a given mass.

In a series of papers
\citep[e.g.,][]{2004MNRAS.348.1078B,2004ApJ...606L..97D,2004ApJ...607L..83M,2004MNRAS.354..111E,2005MNRAS.356.1477D,2005A&A...431..405C,2005astro.ph..9024E}
we have presented results on the properties of galaxy clusters,
extracted from a large--scale simulation and performed with the
{\small GADGET-2} code \citep{SP01.1,2005astro.ph..5010S}, including
the star formation and feedback model of
\cite{2003MNRAS.339..289S,2003MNRAS.339..312S}. This model is
formulated as a sub--resolution model to account for the multiphase
nature of the interstellar medium (ISM), and contains a
phenomenological model of energy feedback from galactic winds
triggered by supernova (SN) explosions (see Section \ref{sec:sims}).
The present paper is specifically aimed at discussing the numerical
robustness of the results we obtained, by analysing an extended set of
re--simulations of galaxy clusters, spanning a fairly large range both
in cluster mass and numerical resolution.  More specifically, we will
primarily address the following two questions: {\em (a)} How does
numerical resolution affect the properties of the diffuse baryons and
the distribution of star formation within clusters?  {\em (b)} How
large is the impact of artificial heating and how can numerical
parameters (i.e., force softening and mass--ratio between gas and DM
particles) be chosen to minimize its effect?

The plan of the paper is as follows. In Section~2, we provide the details of
the simulated clusters. We analyze in Section~3 the effects of changing
resolution over a fairly large range, up to a factor 45 in particle masses. In
this section we will also discuss how the adopted feedback from galactic winds
affects the resolution dependence of measured properties for the simulated
clusters. In Section~4, we discuss the role of different sources of numerical
heating. Finally, we will summarize our results and draw our main conclusions
in Section~5.

\section{The simulations} 
\label{sec:sims}
Our simulations were carried out with {\small GADGET-2}
\citep{2005astro.ph..5010S}, an improved version of the parallel
Tree-SPH simulation code {\small GADGET} \citep{SP01.1}. It uses an
entropy-conserving formulation of SPH \citep{2002MNRAS.333..649S}, and
includes radiative cooling, heating by a uniform redshift--evolving UV
background \citep{1996ApJ...461...20H}, and a treatment of star
formation and feedback processes. The prescription of star formation
is based on a sub--resolution model to account for the multi--phase
nature of the interstellar medium (ISM), where the cold phase of the
ISM is the reservoir of star forming gas \citep[][ SH03
hereafter]{2003MNRAS.339..289S}. As for the feedback, SH03 followed a
phenomenological scheme to include the effect of galactic winds, whose
velocity, $v_w$, scales with the fraction $\eta$ of the SN-II feedback
energy that contributes to the winds, as $v_w\propto \eta^{1/2}$ (see
eq.[28] in SH03). The total energy provided by SN-II is computed by
assuming that they originate from stars with mass $>8\,{\rm M}_\odot$
for a \cite{1955ApJ...121..161S} initial mass function (IMF), with
each SN releasing $10^{51}$ ergs. As discussed in the following, we
will assume $\eta=0.5$ and 1, yielding $v_w\simeq 340$ and 480 km
s$^{-1}$, respectively, while we will also explore the effect of
switching off galactic winds altogether.

We consider two sets of clusters, which have been selected from
different parent cosmological boxes. Initial conditions for both sets
have been generated using the Zoomed Initial Condition (ZIC) technique
by \cite{1997MNRAS.286..865T}. This technique increases the mass
resolution in a suitably chosen high--resolution Lagrangian region
surrounding the structure to be re-simulated. It then adds additional
initial displacements, assigned according to the Zeldovich
approximation \citep[e.g.][]{1989RvMP...61..185S}, from the newly
sampled high--frequency modes which were not assigned in the
low--resolution parent simulation. Furthermore,the mass resolution is
progressively degraded in more distant regions, so as to save
computational resources while still correctly describing the
large--scale tidal field of the cosmological environment.

\begin{table}
\centerline{
\begin{tabular}{lccc}
\hline Cluster run & $M_{\rm vir}$ & $R_{\rm vir}$ & $T_{mw}$\\
\hline
\multicolumn{4}{l}{\tarkin}\\
CL1 & 13.7 &  2.3 &  5.7 \\
CL2 &  2.9 &  1.4 &  2.3 \\
CL3 &  2.4 &  1.3 &  2.1 \\
CL4 &  1.6 &  1.1 &  1.7 \\ ~\\
\multicolumn{4}{l}{\hutt}\\
CL5 & 14.9 &  2.4 &  7.0 \\
CL6 &  1.1 &  1.0 &  1.4 \\
\hline
\end{tabular}
}
\caption{Basic properties of the 6 clusters belonging to the two sets
  of simulations. Column 1: cluster name; Column 2: virial mass,
  defined as the total mass contained within the virial radius (units
  of $10^{14}h^{-1}{\rm M}_\odot$); Column 3: virial radius, defined as the
  radius encompassing an average density equal to the virial density
  predicted for the assumed cosmology (see also text; units of $\hm$);
  Column 4: mass--weighted temperature computed within $R_{\rm vir}$
  (units of keV). All the values reported refer to the reference runs
  of both simulation sets (see text).}
\label{tab:sets}
\end{table}

\begin{table}
\centerline{
\begin{tabular}{lcccc}
\hline Run name & $m_{\rm DM}$ & $m_{\rm gas}$ & $\epsilon_{\rm Pl}$ &
$z_\epsilon$ \\
\hline
\tarkin ~LR & 46.4 & 6.93 & 7.5 & 2\\
\tarkin ~MR & 15.5 & 2.31 & 5.2 & 2\\
\tarkin ~HR &  4.6 & 0.69 & 3.5 & 2\\
\tarkin ~VR &  1.0 & 0.15 & 2.1 & 2\\
\hutt       & 11.3 & 1.69 & 5.0 & 5\\ 
\hline
\end{tabular}
}
\caption{Parameters defining the two sets of runs (see also
  text). Column 1: name of the run; Column 2: mass of the
  high--resolution DM particles (in units of $10^8h^{-1}{\rm M}_\odot$); Column
  3: mass of the gas particles (in units of $10^8h^{-1}{\rm M}_\odot$); Column
  4: Plummer--equivalent gravitational force softening at $z=0$ in the
  high--resolution region (units of $h^{-1}{\rm kpc}$). Column 5: redshift
  of transition from physical to comoving softening.}
\label{tab:res}
\end{table}

Once initial positions and velocities are assigned, a DM--only run is
performed to check for any contamination of the surroundings of the cluster
virial region by heavy particles that may have moved in from the low-- to the
high--resolution region. If required, the shape of the Lagrangian
high--resolution region is optimized by trial and error until any such
contamination around the halo of interest is prevented.  With a typical number
of 3-5 trials, we end up with initial conditions which produce a cluster that,
at $z=0$, is free of contaminants out to (4--6)$\,R_{\rm vir}$. Once initial
conditions are created, we split particles in the high--resolution region into
a DM and a gas component, whose mass ratio is set to reproduce the assumed
cosmic baryon fraction. Instead of placing them on top of each other, we
displace gas and DM particles such that the centre of mass of each parent
particle is preserved and the final gas and dark matter particle distributions
are interleaved by one mean particle spacing.

\subsection{The \tarkin}
This set includes four clusters, resimulated at different resolutions,
with virial mass in the range $M_{\rm
  vir}=(1.6$--$13)\times 10^{14}h^{-1}{\rm M}_\odot$ (CL1 to CL4 in Table
\ref{tab:sets}).  
These clusters have been extracted from the 
cosmological hydrodynamical simulation presented by
\cite{2004MNRAS.348.1078B}.
The simulation followed $480^3$ DM
particles and an initially equal number of gas particles, within a box
of $192\,h^{-1}$Mpc on a side, for a flat $\Lambda$CDM model with
$\Omega_m=0.3$, $h=0.7$, $\sigma_8=0.8$ and $\Omega_{\rm
b}=0.04$. With these choices, the masses of the DM and gas
particles are $m_{\rm DM}\simeq 4.6\times 10^9\msun$ and $m_{\rm
gas}\simeq 6.9\times 10^8\msun$, respectively. The force accuracy is
set by $\epsilon_{\rm Pl}=7.5\hk$ for the Plummer--equivalent
softening parameter, fixed in physical units from
$z=0$ to $z=2$, and kept fixed in comoving units at higher
redshifts.

Since initial conditions of the parent simulations have been generated on a
grid, a grid is also used to assign initial displacements for these
resimulations. Initial conditions for each cluster are generated at four
different mass resolutions, corresponding to the basic resolution of the
parent box (low resolution, LR), and to 3 times (medium resolution, MR), 10
times (high resolution, HR) and 45 times (very high resolution, VR) smaller
particle masses. The VR run is not carried out for the CL1 cluster, whose
large mass would result in a too large computational cost. The gravitational
softening in the high--resolution regions is rescaled with the mass of the
particles according to $\epsilon_{\rm Pl}\propto m^{1/3}$, where the LR runs
were set to have the same softening as used in the parent simulation. In Table
\ref{tab:res}, we list the masses of the DM and gas particles, as well as the
force softening for the different resolutions.  At the highest achieved
resolution, each cluster is resolved with at least 1,5 million DM particles
within the virial radius, with CL2 reaching 2,2 million particles.  The
reference runs for the clusters of \tarkin~ have been performed by assuming
that 100 per cent of the energy provided by SNe is carried by winds.  This
gives a wind speed of $v_w\simeq 480\vel$.

\begin{table*}
\centerline{
\begin{tabular}{lll}
\hline Run name & Description & Applied to:\\
\hline
NW  & No winds: winds switched off & CL2-LR,HR; CL4-LR,MR,HR,VR\\
SW  & Strong winds: wind velocity for the \hutt~ at the
reference value of the \tarkin & CL5, CL6 \\
1o8 & Eight times heavier gas particles & CL4-VR\\
S08 & SW runs for clusters of the \hutt~ with $\sigma_8=0.8$ & CL5, CL6\\
\hline
\end{tabular}
}
\caption{Description of the name extensions used for the identification of
  the runs.
}
\label{tab:tests}
\end{table*}

\subsection{The \hutt}
\label{s:hutt}
This set includes 2 clusters having mass $\simeq 10^{15}\msun$ and
$\simeq 10^{14}\msun$ (CL5 and CL6 in Table \ref{tab:sets},
respectively). They have been selected from a larger sample of 20
clusters with masses in the range $5\times 10^{13}-2.3\times
10^{15}\msun$, which have been identified in simulations of
9~Lagrangian regions (Dolag et al., in preparation).  These systems
were extracted from a DM--only simulation with a box size of
$479\,h^{-1}$Mpc of a flat $\Lambda$CDM model with $\Omega_m=0.3$,
$h=0.7$, $\sigma_8=0.9$ and $\Omega_{\rm b}=0.04$
\citep{2001MNRAS.328..669Y}. Differently from \tarkin, the initial
displacements were generated using a `glass' \citep{WH96.1b} for the
Lagrangian particle distribution.  Only one mass resolution was
simulated in this case, corresponding to $m_{\rm DM}=1.13\times
10^9\,h^{-1}{\rm M}_\odot$ and $m_{\rm gas}=1.7\times 10^8\,h^{-1}{\rm
  M}_\odot$ for dark matter and gas within the high--resolution
region, respectively.  As such, this mass resolution is about 4 times
better than the LR runs, and therefore lies intermediate between the
MR and the HR runs of \tarkin.  The softening length is set to
$\epsilon_{\rm Pl}=5.0\, h^{-1}$kpc, fixed in physical units below
$z=5$, while it is kept fixed in comoving units at higher redshift.
%at $z=0$, and is hence very close
%to that used for \tarkin, once rescaled according to mass resolution,
The reference runs for clusters of \hutt~ assume that 50 per cent of
the energy provided by SN (i.e. half of that used for the \tarkin) is
carried by winds. This gives a wind speed of $v_w\simeq 340\vel$.
Unless otherwise stated, our analysis presented in this paper is
restricted to the CL5 and CL6 simulations out of the 20 clusters of
the full \hutt.  We note that for all of our simulations, we set the
smallest allowed value for the SPH smoothing length to $\epsilon_{\rm
  Pl}/4$.

\begin{figure*}
\centerline{
\vbox{
\hbox{
\psfig{file=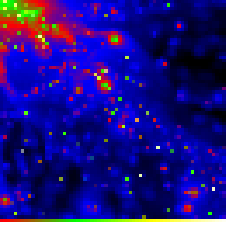,width=7.5cm} 
\psfig{file=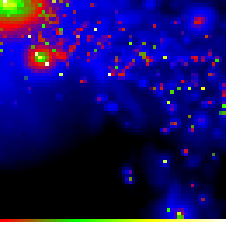,width=7.5cm} 
}
\hbox{
\psfig{file=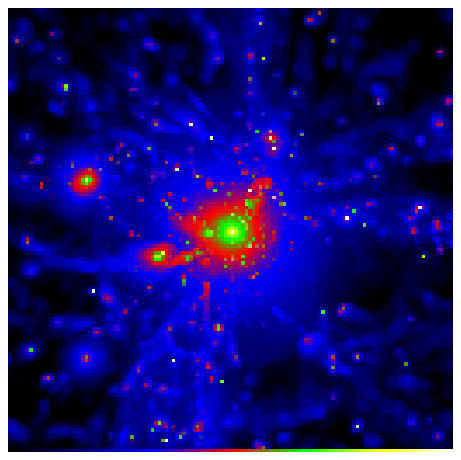,width=7.5cm} 
\psfig{file=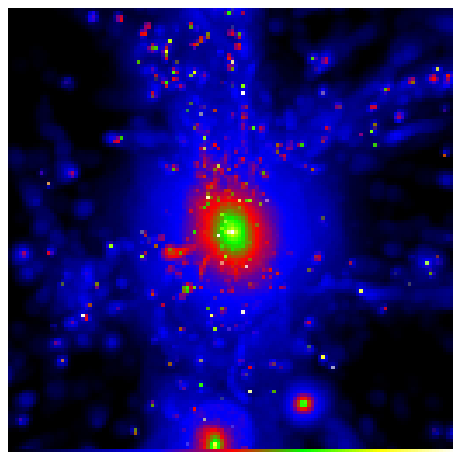,width=7.5cm} 
}
\hbox{
\psfig{file=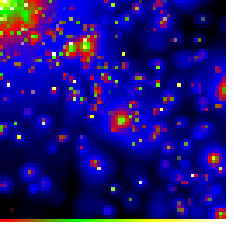,width=7.5cm} 
\psfig{file=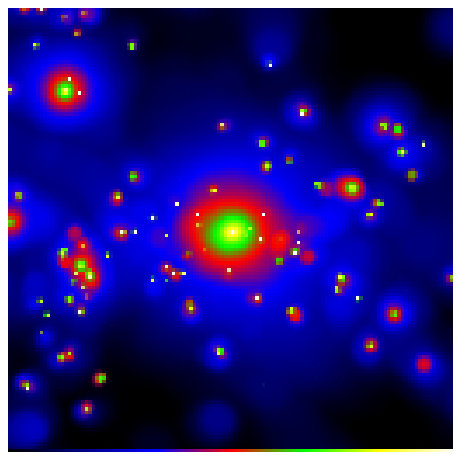,width=7.5cm} 
}
}}
\caption{Maps of the gas density for the six simulated clusters,
  in a region encompassing two virial radii around each cluster centre.
  Top--left to bottom--right panels are for the CL1 to CL6 clusters. For the
  clusters belonging to \tarkin\ (CL1 to CL4), we show maps for the highest
  resolution runs (HR for CL1 and VR for CL2 to CL4). In all cases, the maps
  refer to the runs with strong winds (i.e.~$v_w\simeq 480\vel$). The small
  bright knots mark the ``galaxies'', i.e.~the places where high density gas
  is undergoing cooling and star formation. }
\label{fi:maps} 
\end{figure*}

\vspace{0.3truecm} In summary, our two sets of simulated clusters
differ in the following aspects:
\begin{description}
\item[(a)] The value of $\sigma_8$ changes from 0.8 to 0.9 when going
  from \tarkin\ to \hutt. To explicitly check for the effect of
  changing the normalization $\sigma_8$, we will also analyse
  additional simulations of the CL5 and CL6 clusters of \hutt~ using
  $\sigma_8=0.8$.
\item[(b)] The initial Lagrangian positions of DM and gas particles
  were taken to be a grid for \tarkin ~and a glass for \hutt. We will
  also present results for CL5 of \hutt~ using grid initial
  conditions, to check explicitly for effects due to a grid vs. a
  glass setup.
\item[(c)] The velocity of the galactic winds is $v_w\simeq 480\vel$
  and $340\vel$ for \tarkin ~and \hutt, respectively. To address this
  important difference, we have rerun the CL5 and CL6 clusters from
  \hutt~ with $\sigma_8=0.8$ and with the stronger feedback, to make
  them fully comparable with the clusters of \tarkin.  
\item[(d)] The transition from physical to comoving softening takes
  place at $z=5$ for \hutt ~and at $z=2$ for \tarkin. To test
  the effect of a change of this transition redshift we have rerun
  CL6 of \hutt ~with strong winds, also using the same choice of softening
  as for \tarkin.
\end{description}

We will use the convention that a certain simulation of each cluster
is labeled by the name of the cluster itself, followed, when required,
by a label which specifies the resolution used. In this way, CL1-MR
will indicate the reference run of medium--resolution of the CL1
cluster from \tarkin, while CL5 will designate the reference run of
this cluster from \hutt. For the latter, we do not specify the
resolution, since clusters from \hutt~ are simulated at only one
resolution.  In addition, a further extension of the name of each run
will be provided whenever the run differs from the reference run of
the set it belongs to. The list of such extensions and their
description is given in Table \ref{tab:tests}. For instance, CL4-HR
indicates the high--resolution run of the CL4 cluster, using the
standard setup of \tarkin, while CL4-HR-NW stands for the same
simulation, but neglecting the effect of galactic winds.

We show in Figure \ref{fi:maps} the gas density maps of the six
simulated clusters within boxes each having a size of $4R_{\rm
  vir}$. The four clusters from the \tarkin~ are shown in their
highest resolution version. 

\subsection{Basics of the simulation analysis}

In the first step of our simulation analysis we determine suitable cluster
centers. These are defined as the position of the most bound particle among
those grouped together by a FOF algorithm with linking length $b=0.15$ (in
units of the mean interparticle separation in the high--resolution region).
Once the center is identified, we apply a spherical overdensity algorithm to
determine the virial radius, \rvir. We here defined this as the radius that
encompasses an average density equal to the virial density for the adopted
cosmological model, $\rho_{\rm vir}(z)=\Delta_c(z)\rho_{c}(z)$
($\rho_c(z)=[H(z)/H_0]^2\rho_{c,0}$ is the critical density at redshift $z$),
where the overdensity $\Delta_c(z)$ is computed as described by
\cite{1996MNRAS.282..263E}, with $\Delta_c(0)\simeq 100$ for the
assumed cosmology. The virial mass, \mvir, is simply the mass
contained within the virial radius. In Table~\ref{tab:sets}, we provide the
typical values of \rvir, \mvir\ and mass--weighted temperature, \tmw, for the
six simulated clusters. Profiles of gas-related quantities are computed
within 200 equispaced linear radial bins, out to 2\rvir, starting from a
minimum radius which contains 100 gas particles. As shown by
\cite{2002MNRAS.336..409B}, numerically stable results can be expected for
this choice in non--radiative simulations of galaxy clusters.

\begin{figure*}
\centerline{
\hbox{
\psfig{file=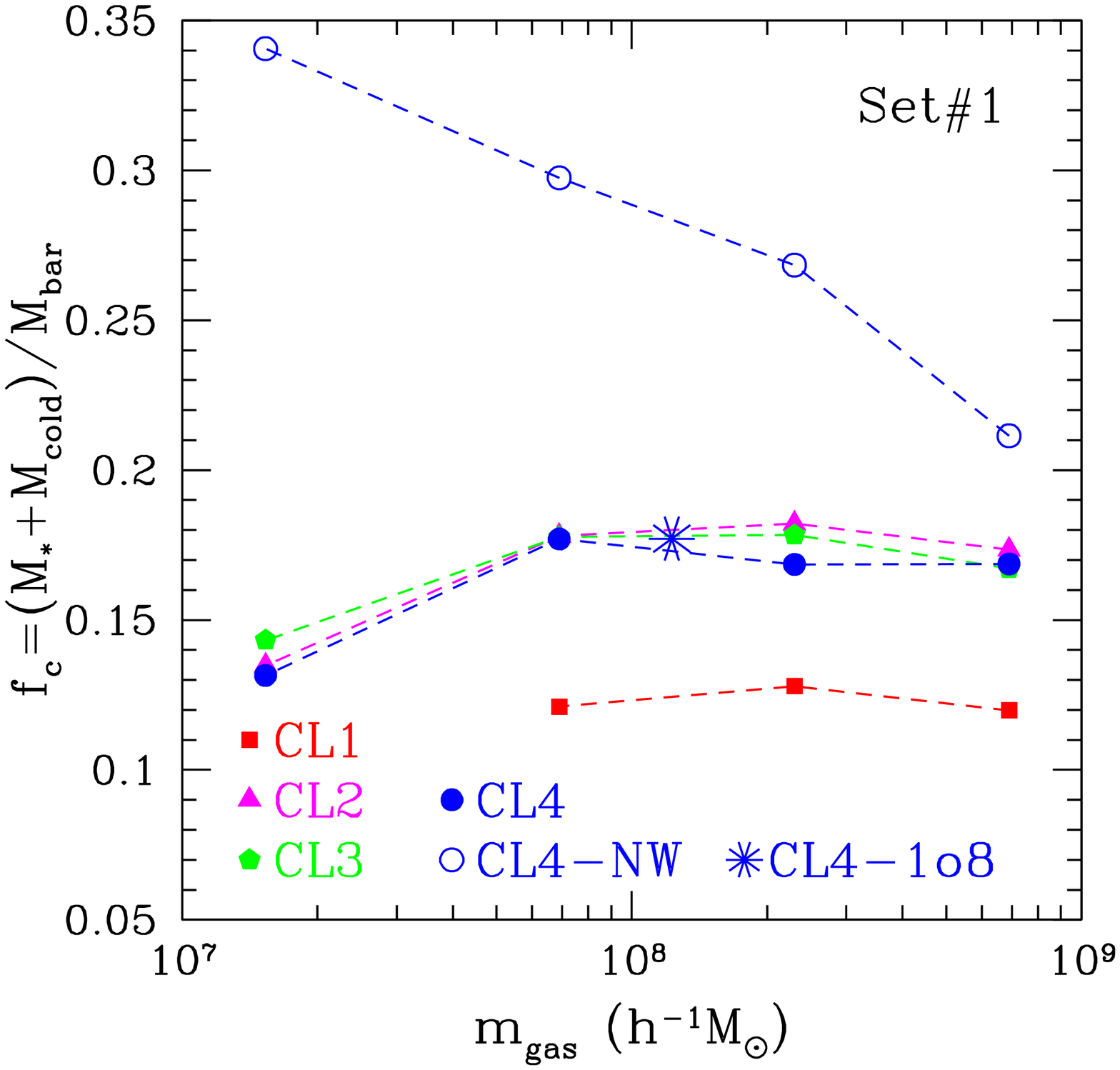,width=8.cm} 
\psfig{file=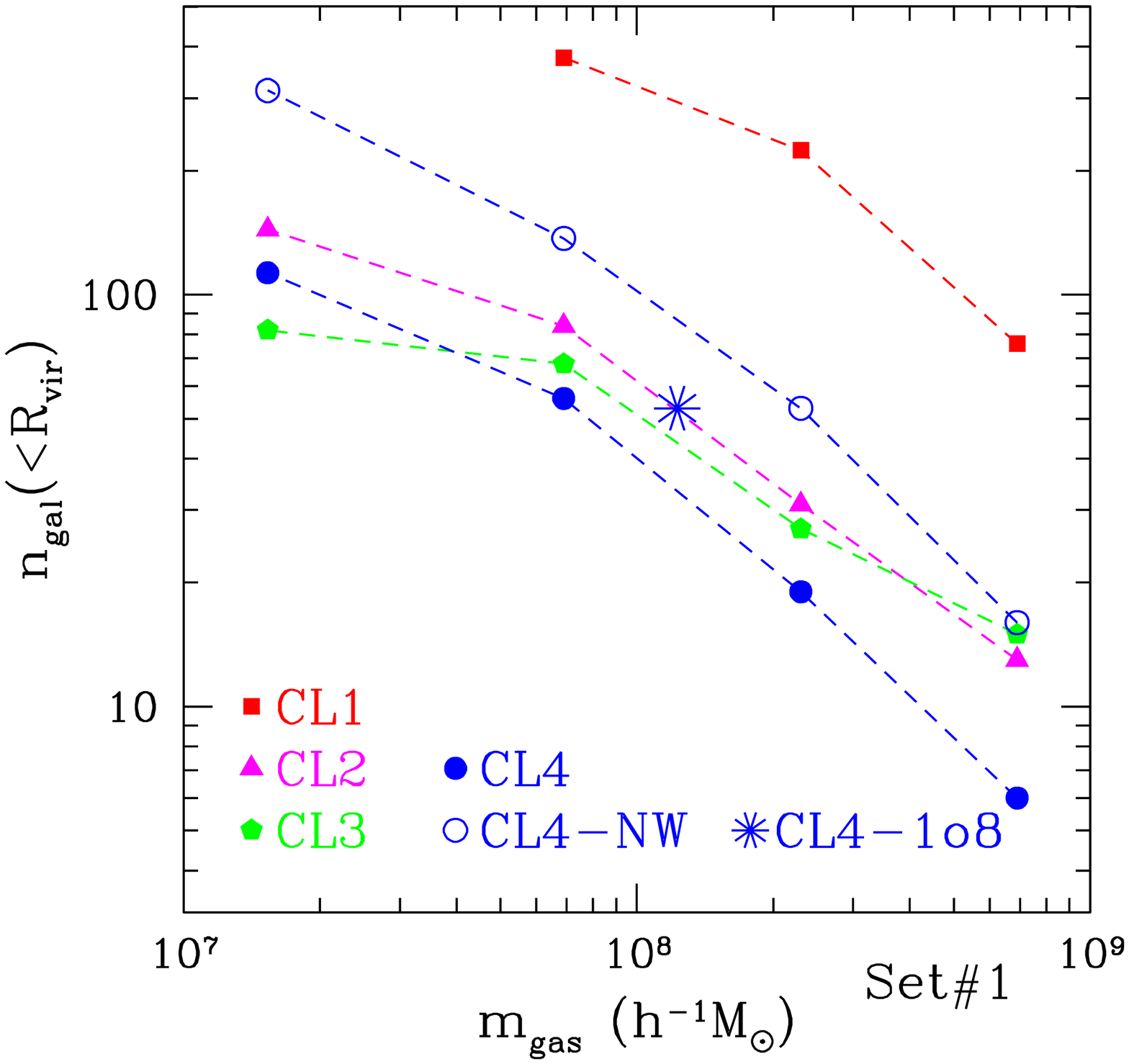,width=8.cm} 
}
}
\caption{The fraction of cooled baryons $f_c$ (left panel), and the
  number of galaxies within $R_{\rm vir}$, as functions of the mass of
  the gas particle, for the clusters of \tarkin\ at different
  resolution. Filled symbols are for the ``reference'' runs. The open
  circles are for the runs of the CL4 cluster with wind feedback
  turned off. The asterisk is for the CL4 run at very high resolution
  using 8 times heavier gas particles (CL4-1o8), so that the gas particle
  mass is similar to that of the DM particles in the high--resolution
  region.}
\label{fi:fstar_res}
\end{figure*}

We identify galaxies in our simulations by applying the SKID
algorithm\footnote{See {\tt
    http://www-hpcc.astro.washington.edu/tools/skid.html} }
\citep{2001PhDT........21S} to the distribution of star particles. In the
following, we provide a short description of our algorithm, while a more
detailed discussion and presentation of tests is provided elsewhere (Murante
et al. 2005, in preparation). Briefly, the SKID algorithm works as follows:
\begin{itemize}
\item An overall density field is computed by using the distribution of all
  the particle species, i.e. DM, gas and star particles.  The density is
  estimated with a SPH spline--kernel, using a given number $N_{sm}$  of
  neighbour particles.
\item The star particles are moved along the gradient of the density
  field in steps of $\tau/2$. When a particle begins to oscillate
  inside a sphere of radius $\tau/2$, it is stopped. In this way,
  $\tau$ can be interpreted as the typical size of the smallest
  resolved structure in the distribution of the star particles.
\item When all particles have been moved, they are grouped using a
  friends-of-friends (FOF) algorithm applied to the moved particle
  positions. The linking length is again $\tau/2$.
\item The binding energy of each group identified in this way is computed by
  accounting for all the particles inside a sphere centered on the center of
  mass of the group and having radius $2\tau$ (for the moved particles, their
  initial positions are used in the computation of the potential). This
  binding energy is then used to remove from the group all the star particles
  which are recognized as unbound. Finally, we retain 
such a SKID--group of stars as a
  galaxy if it contains at least 32 particles after the removal of unbound
  stars.
\end{itemize}

The resulting list of objects identified by SKID depends on the choice of two
parameters, namely $\tau$ and $N_{sm}$. After many experiments, and resorting
to visual inspection in all cases, we found that a complete detection of bound
stellar objects requires use of a set of different values of $N_{sm}$. We used
$N_{sm}=16,32,64$, and define a ``galaxy'' to be the set of star particles
which belong to a SKID group with any one of the above $N_{sm}$ values. If a
star particle belongs to a SKID group for one value of $N_{sm}$ and to another
group for a different $N_{sm}$, then the groups are ``joined'' and are
considered as forming a single galaxy. All star particles not linked to any
galaxy are considered to be part of a diffuse stellar component in the cluster
\citep{2004ApJ...607L..83M}. As for $\tau$, since it
roughly corresponds to the size of the smallest resolved structure, we adopt
$\tau=2.8\epsilon_{\rm Pl}$, which is the scale where the softened force
becomes equal to the Newtonian force.

\section{The effect of mass resolution}

Checking the stability of simulation results against numerical
resolution is always of paramount importance to assess their
robustness.  For hydrodynamical simulations of clusters that only
account for non--radiative physics, it has been shown that numerically
reliable estimates of global cluster properties, such as temperature
and X--ray luminosity, are obtained when each halo is resolved with a
few tens of thousands of gas particles within the virial region, a
resolution which is also enough to result in converged radial profiles
for the gas density, temperature and entropy down to a few percent of
the virial radius \citep[e.g., ][ and references
therein]{1995MNRAS.275..720N,2002MNRAS.336..409B}. However, the
situation is considerably less clear when additional physical
processes are introduced.  For example, the efficiency of radiative
cooling in simulations is known to sensitively depend on the adopted
mass resolution. At the same time, the effects of feedback from
galactic winds may also depend on numerical resolution, and it is
unclear whether this resolution dependence favourably counteracts the
increasing efficiency of cooling.  We generated the initial conditions
for \tarkin~ at four different resolutions precisely for the purpose of
checking in detail the resolution--dependence of the non-trivial
physical processes included in our simulations.

\begin{figure*}
\centerline{
\hbox{
\psfig{file=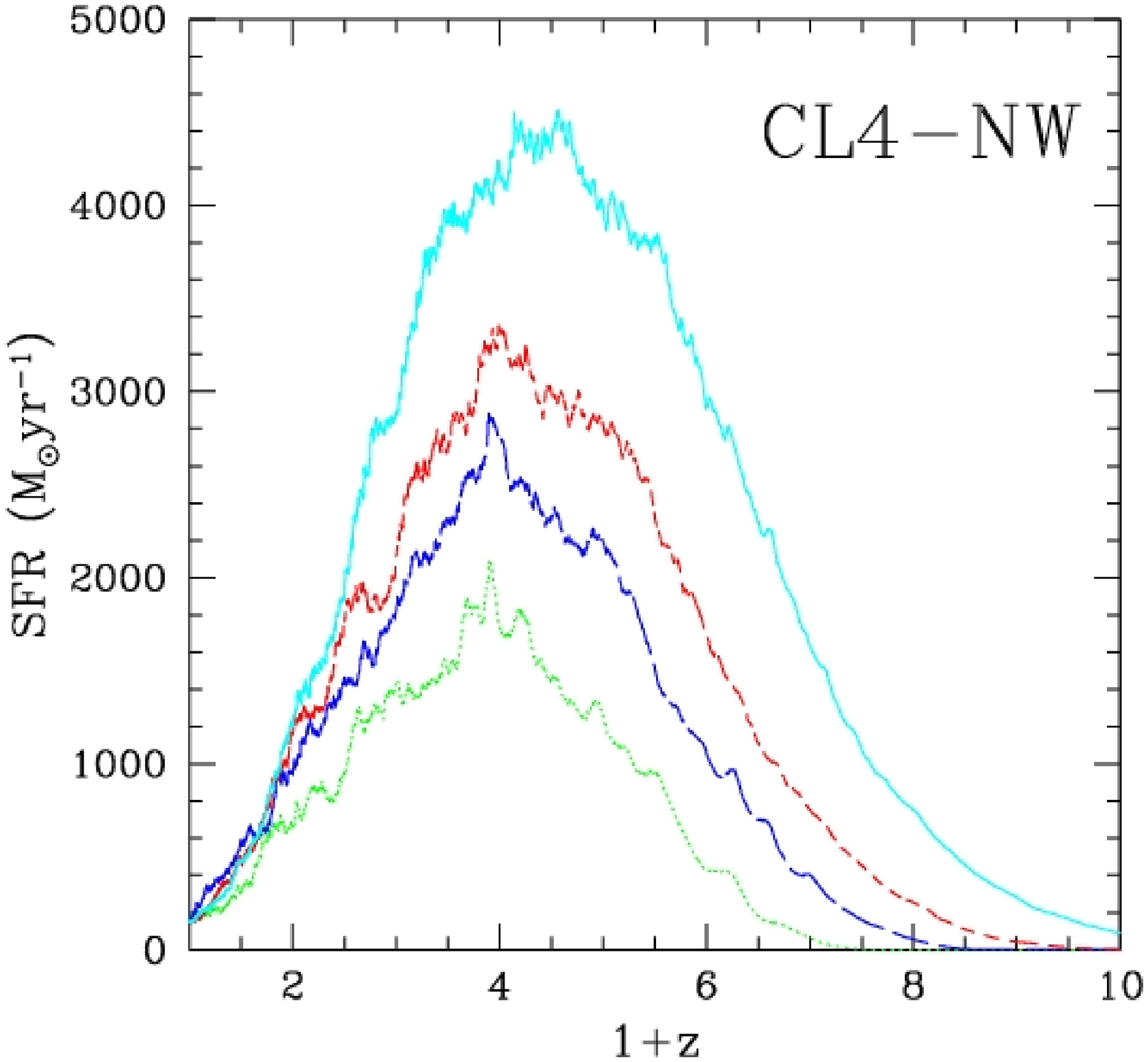,width=9.cm} 
\psfig{file=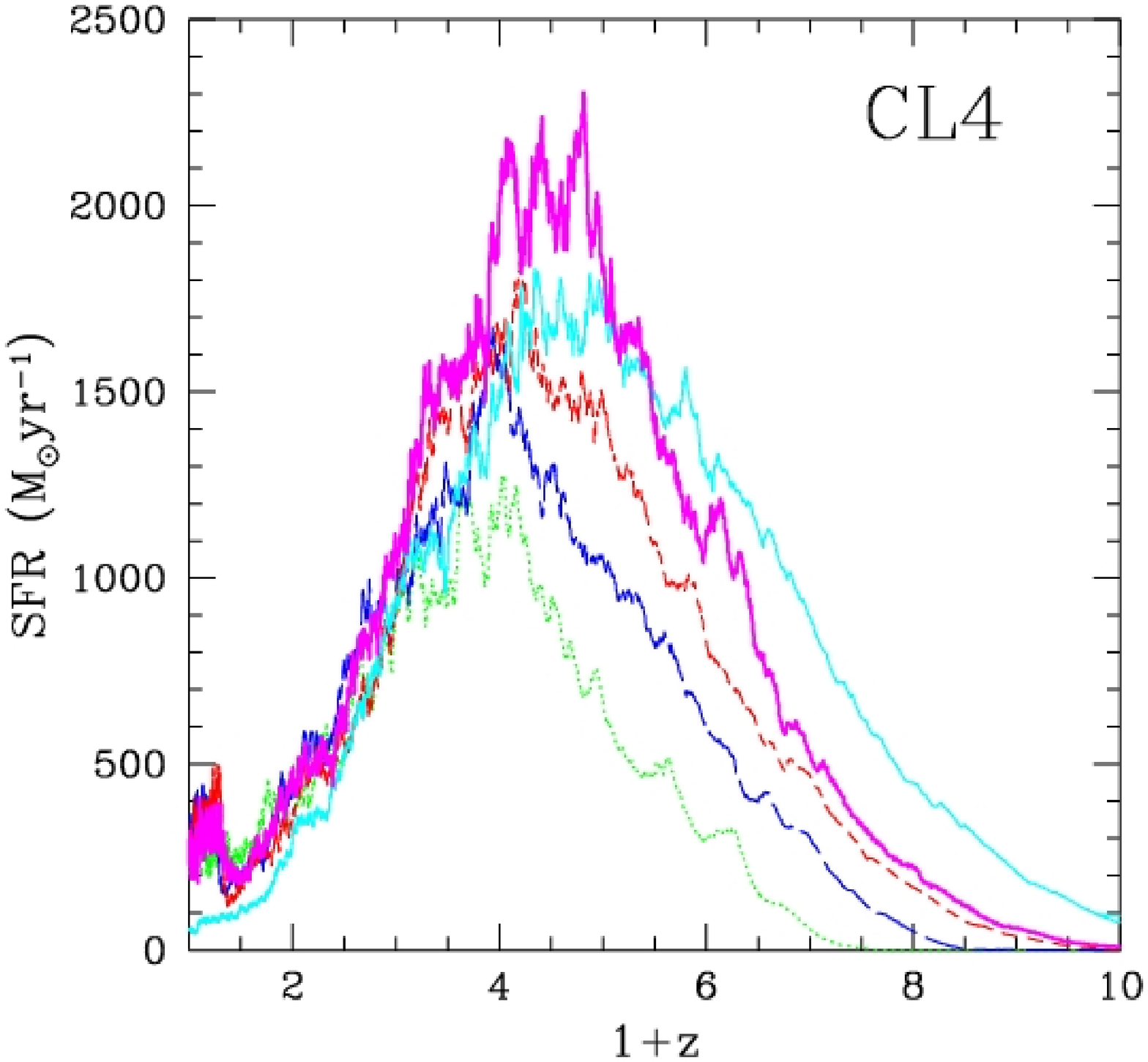,width=9.cm} 
}}
\caption{The star formation rate as a function of the mass resolution
  for the CL4 cluster, both excluding (left panel) and including
  (right panel) the effect of galactic winds. In both panels, solid,
  short--dashed, long--dashed, and dotted curves correspond to the VR,
  HR, MR and LR runs, respectively. In the right panel, the heavy
  solid curve is for the VR run, where gas particles have nearly the
  same mass as the DM particles (1o8 run in
  Table~\ref{tab:tests}). Note the different scales on the vertical axes in
  the two panels.}
  \label{fi:sfr_profs_res} 
\end{figure*}

\subsection{Star formation}

In Figure~\ref{fi:fstar_res}, we show the dependence of the fraction
of cooled baryons and of the number of identified galaxies within
\rvir ~on the mass of the gas particles. In the absence of feedback by
galactic winds, the CL4 runs develop the typical runaway of cooling as
a function of the mass resolution.  The fraction of cooled baryons
steadily increase from 21 per cent at the lowest resolution to 34 per
cent at the highest achieved resolution, with no indication of
convergence. A part of the origin of this runaway is illustrated in
the left panel of Figure \ref{fi:sfr_profs_res}, where we show the
star formation history for the CL4 runs at the different resolutions,
in the absence of winds. Increasing the resolution leads to an earlier
onset of star formation within ever smaller first--collapsing halos,
and at low redshift, cooling and star formation appear to never enter
in any self--regulated regime. As a result, the star formation
efficiency progressively increases with resolution at all epochs.

On the other hand, including feedback by galactic winds is quite
effective in regulating the process of star formation. In this case,
there is no evidence for a systematic increase of the fraction of
cooled baryons with increasing resolution. In fact, the three poor
clusters have rather similar values of $f_c\sim 15$ per cent, and the
massive cluster has a value of $f_c\simeq 12$ per cent. This result
demonstrates that a converged estimate of the star fraction is
obtained already at modest resolution with our feedback scheme.  This
also extends the earlier results by \cite{2003MNRAS.339..312S} to the
scale of clusters.  Intriguingly, we note that all the runs at very
high resolution (VR) even show a small but systematic decrease of the
cold gas fraction.

Including feedback from winds has the twofold effect of suppressing
the resolution--dependence of the cumulative efficiency and to make
the SFR history almost independent of resolution at $z\mincir 2$--3
(see the right panel of Fig.~\ref{fi:sfr_profs_res}). The enhanced
star formation activity at high--redshift of the high resolution runs
produces more efficient gas pre--heating, which has the effect of
inhibiting star formation at later times.  This presumably explains
that the star formation rate of the VR run at $z\mincir 2$ is even
below those of the lower--resolution runs. Assuming that the
smallest resolved halos where cooling can take place contain $\sim$100
DM particles, we find their escape velocity to be of  order
100$\vel$ for the low--resolution (LR) runs, and a factor of about
three smaller for the VR runs of  \tarkin. Therefore, the wind
velocity is always larger than the escape valocity of the smallest
halos which are first resolved at high redshift, thereby implying that
our feedback scheme is efficient in preventing star
formation already in the first generation of resolved galaxies.

An alternative explanation for the decrease of the cold fraction
in the VR runs could be that the suppression of star formation in the
highest resolution run is related to some numerical effect. For
instance, one may argue that an improved resolution provides a more
accurate description of the gas behaviour at cooling
interface. Indeed, a coarse description of this interface is expected
to cause spurious gas cooling \citep{Pearce1999ApJ...521L..99P}, a
feature that should however be weak in the entropy conserving
formulation of SPH \citep{2002MNRAS.333..649S} implemented in our
code. However, if this was really the case, one would expect the same
effect to appear also in runs without galactic winds (NW runs). But
the left panel of Fig.~\ref{fi:sfr_profs_res} clearly demonstrates
that this is not the case.  Hence, the stable behaviour of star
formation with increasing resolution is more likely related to the
inhibiting effect of more efficient high--$z$ feedback onto later
generations of galaxies.

\begin{figure}
\centerline{
\psfig{file=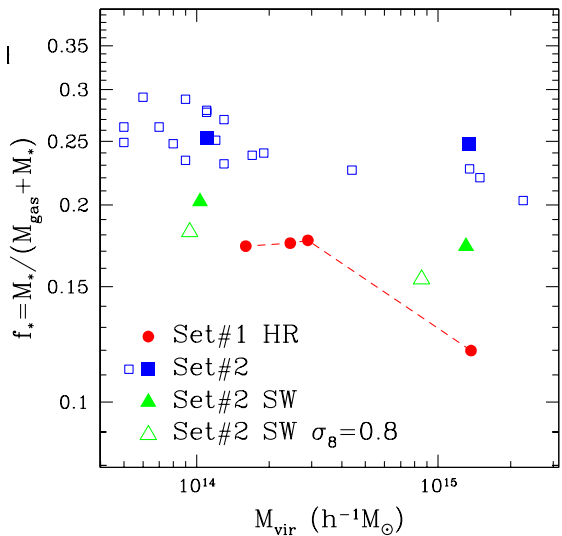,width=8.cm} 
}
\caption{The stellar fraction, $f_*$, as a function of the cluster virial
  mass, for the different simulation sets. Filled circles connected by a
  dashed line are for the high--resolution version of the clusters of \tarkin.
  Squares are for all the clusters of \hutt, with the two filled squares
  showing the results for simulations CL5 and CL6. Filled triangles are for
  CL5 and CL6, but using the same velocity of galactic winds, $v_w\simeq
  480\vel$ as for \tarkin. The open triangles are the same simulations as the
  filled triangles, but using the power spectrum normalization of \tarkin,
  $\sigma_8=0.8$. Therefore, filled circles and open triangles come from
  different sets of initial conditions, which however have been evolved using
  an identical simulation set-up.}
\label{fi:fstar_mvir}
\end{figure}

\begin{figure*}
\centerline{
\hbox{
\psfig{file=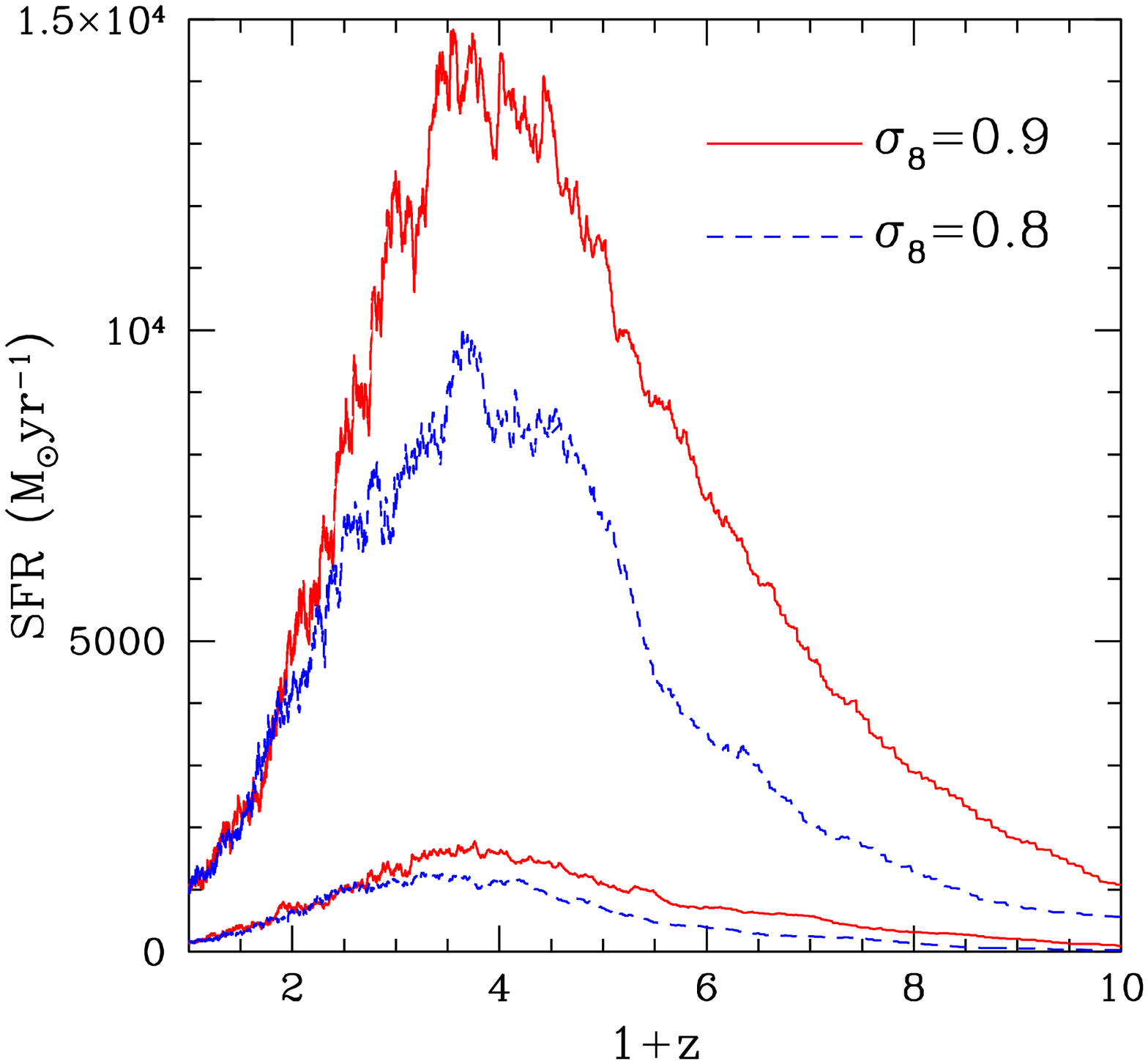,height=8.cm} 
\psfig{file=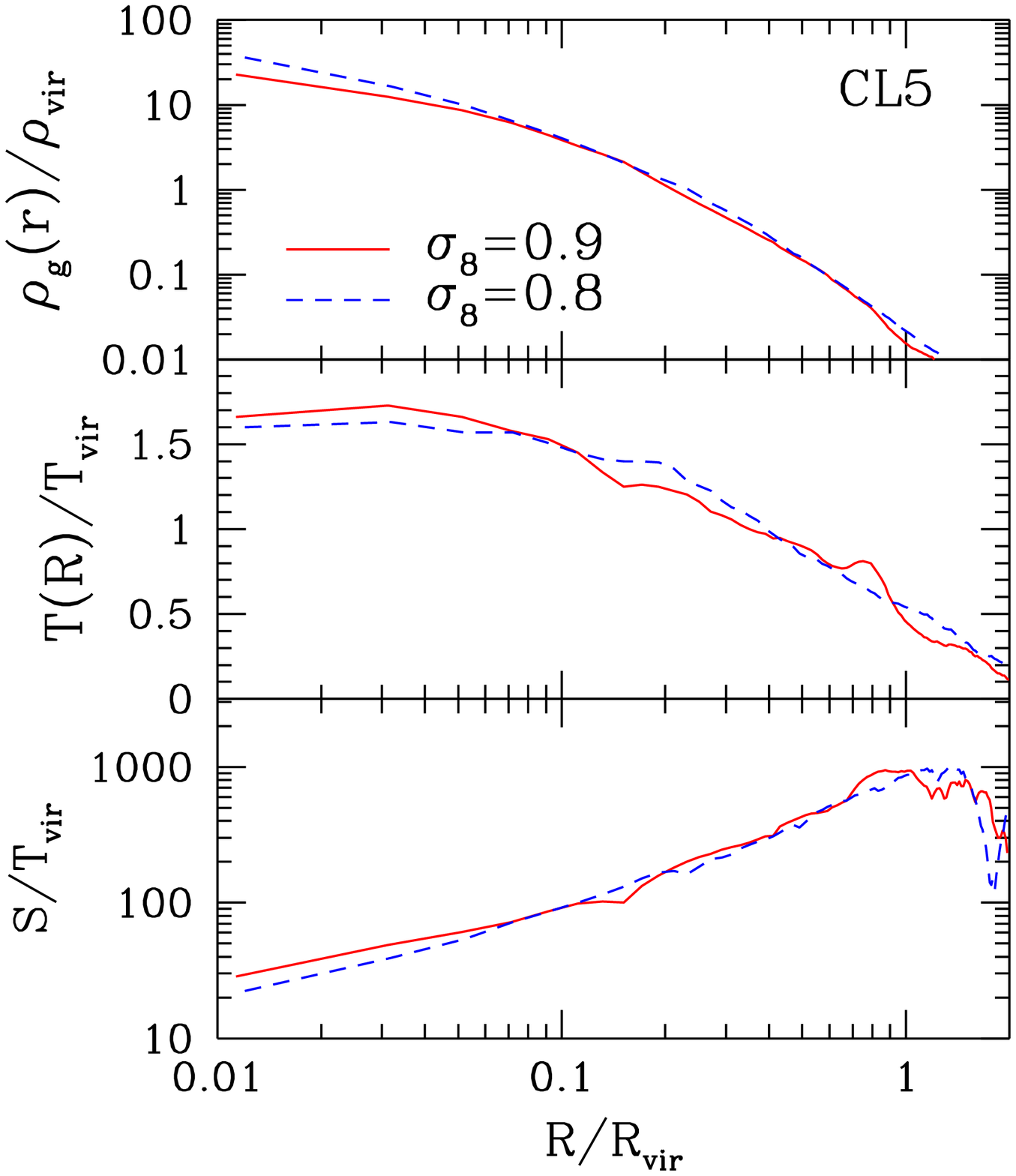,height=8.cm} 
}}
\caption{The effect of changing the power spectrum normalization
  $\sigma_8$. Left panel: the effect on the star--formation history for the
  CL5 and CL6 clusters (upper and lower pairs of curves), both simulated with
  the strong--wind (SW) feedback; solid and dashed curves refer to
  $\sigma_8=0.9$ and $\sigma_8=0.8$, respectively. Right panel: the effect on
  gas--related profiles for the CL5 cluster.}
 \label{fi:sig8}
\end{figure*}

The number of identified galaxies within each cluster increases with better
resolution and with the cluster mass, as expected. As shown in the right panel
of Figure~\ref{fi:fstar_res}, the number of identified {\em bona--fide}
galaxies grows by more than one order of magnitude when passing from the LR to
the VR runs. In line with the behaviour of the stellar fraction, the number of
galaxies for the CL4 cluster increases by a factor of about three when the
wind feedback is switched off, almost independent of the resolution. We defer
a detailed analysis of the stellar mass function of the identified galaxies
and of the diffuse stellar component as a function of resolution to
forthcoming work (Murante et al., in preparation).

In Figure \ref{fi:fstar_mvir}, we summarize the results on the star fraction
by plotting it as a function of the cluster virial mass, for the reference
runs of both \tarkin~ (filled circles) and of all the 20 clusters of \hutt~
(squares). Both simulation sets confirm a trend of decreasing star fraction as
a function of the cluster mass. However, clusters belonging to \hutt~ show a
stellar fraction which is systematically higher than that of \tarkin, the
difference being larger than any possible object-to-object intrinsic scatter
induced by the varying dynamical histories of different clusters. However, we
recall that the two sets of cluster simulations differ in the strength of the
adopted feedback and in the normalization of the power spectrum.

To investigate the effect of the different feedback, we have run additional
simulations of CL5 and CL6 of \hutt~ by increasing the wind speed to the same
value as used for \tarkin~ (SW runs). The results for the stellar fraction are
shown with filled triangles in Fig.~\ref{fi:fstar_mvir}. Although increasing
the feedback efficiency produces a significant suppression of $f_*$, the
effect is still not large enough to fully account for the difference between
the two simulation sets, thus suggesting that the residual difference is due
to the different $\sigma_8$ values.

We have explicitly verified this by repeating the CL5--SW and CL6--SW
runs by also decreasing $\sigma_8$ to 0.8.  The comparison of the
resulting star formation histories is shown in the left panel of
Figure~\ref{fi:sig8}. Reducing $\sigma_8$ results in a significant
change in the timing of structure formation and, correspondingly, to a
delay in the high--redshift star formation, with a suppression of its
peak at $z\sim 3$.  The resulting stellar fraction further decreases
from $f_*= 17.5$ per cent to 15.6 per cent for CL5, and from $f_*=
20.3$ per cent to 18.1 per cent for CL6.  As shown in
Fig.~\ref{fi:fstar_mvir}, the reduction of $f_*$ connected to the
power spectrum amplitude finally brings the values of the cooled
fraction for the simulated clusters of \tarkin~ and \hutt\ into good
agreement.

The decreasing trend of the star fraction with cluster mass, shown in
Fig.~\ref{fi:fstar_mvir}, is in qualitative agreement with
observational results \citep[][ cf. also
\citealt{2001MNRAS.326.1228B}]{2003ApJ...591..749L,2004AJ....128.1078R}.
\cite{2003ApJ...591..749L} used K--band data from the 2MASS survey to
trace the stellar population in clusters, and ROSAT--PSPC data to
measure the corresponding gas mass. They found $f_*\simeq 10$ per cent
for clusters with mass of about $10^{15}\msun$, increasing to $\simeq
15$ per cent for clusters with $10^{14}\msun$. These values are not
far from those obtained from our runs with stronger feedback and
$\sigma_8=0.8$. Also, it is worth pointing out that the observational
census of the cluster stellar population by \cite{2003ApJ...591..749L}
does not include diffuse stars, whose contribution may not be
negligible both for real \citep[e.g., ][, for a
review]{2004IAUS..217...54A} and simulated
\citep{2004ApJ...607L..83M,2004MNRAS.355..159W,2005MNRAS.357..478S}
clusters.  Note that our estimates of $f_*$ include all stars found
inside the simulated clusters.

\begin{figure*}
\centerline{
\vbox{
\hbox{
\psfig{file=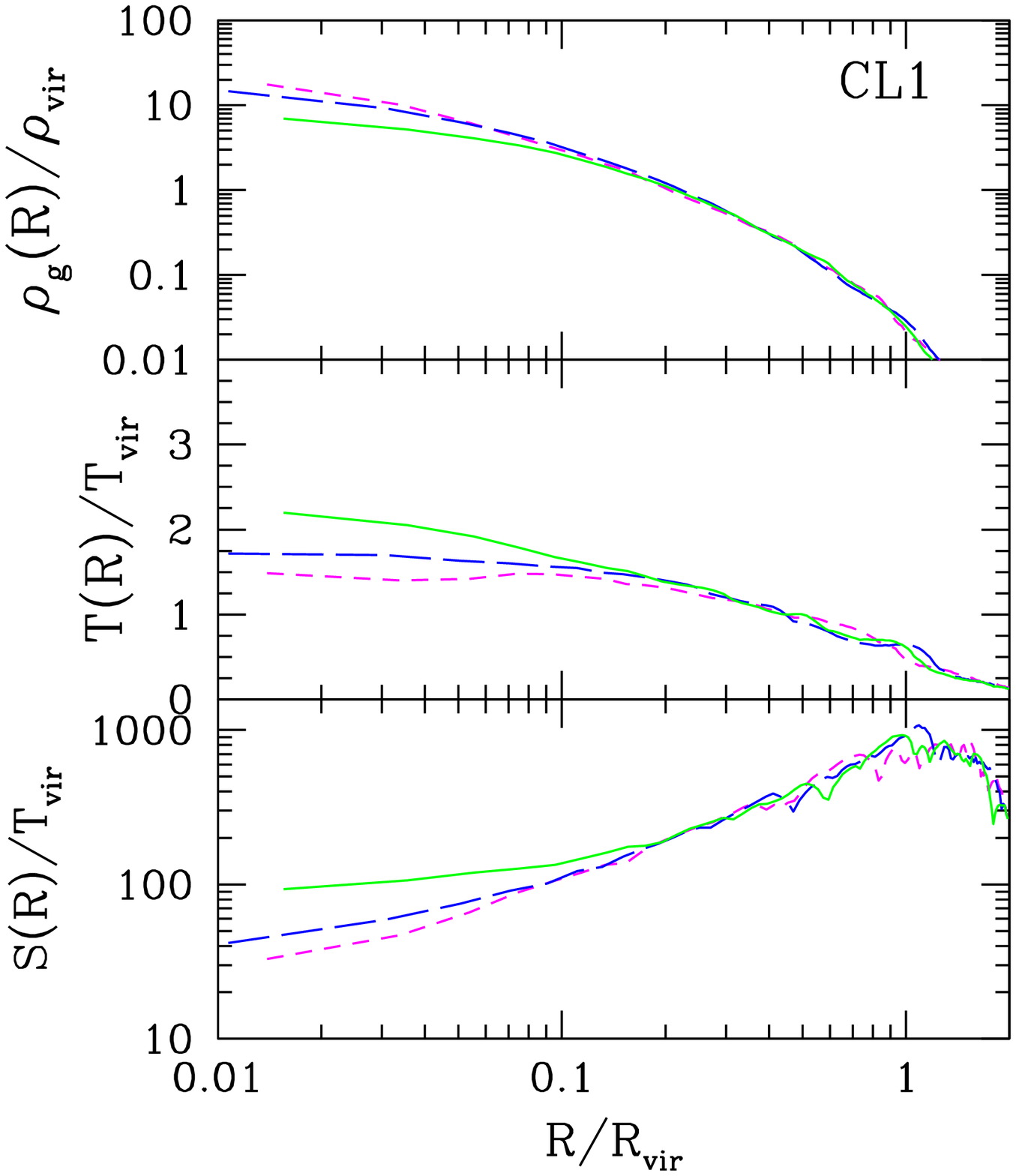,width=7.5cm} 
\psfig{file=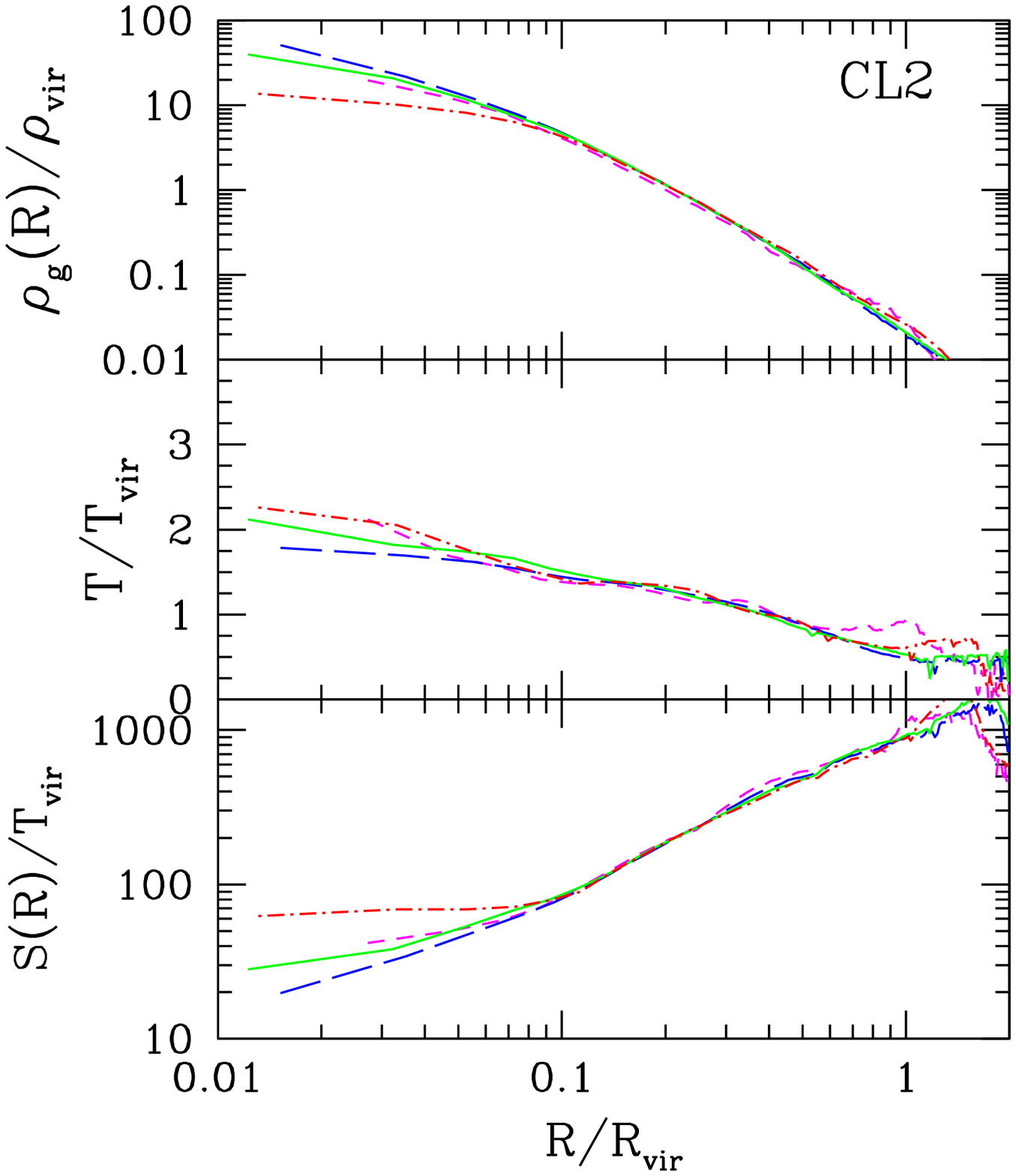,width=7.5cm} 
}
\hbox{
\psfig{file=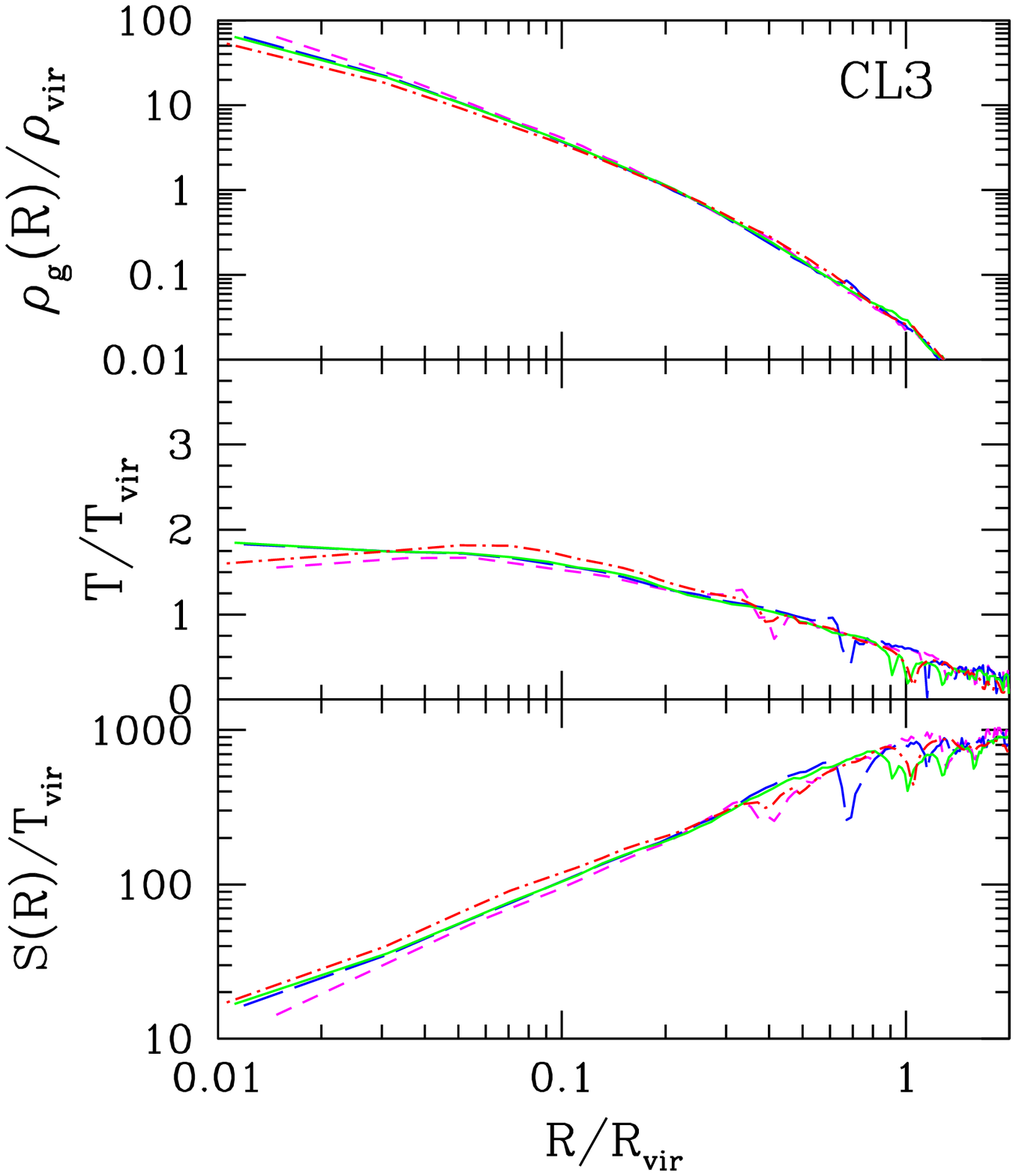,width=7.5cm} 
\psfig{file=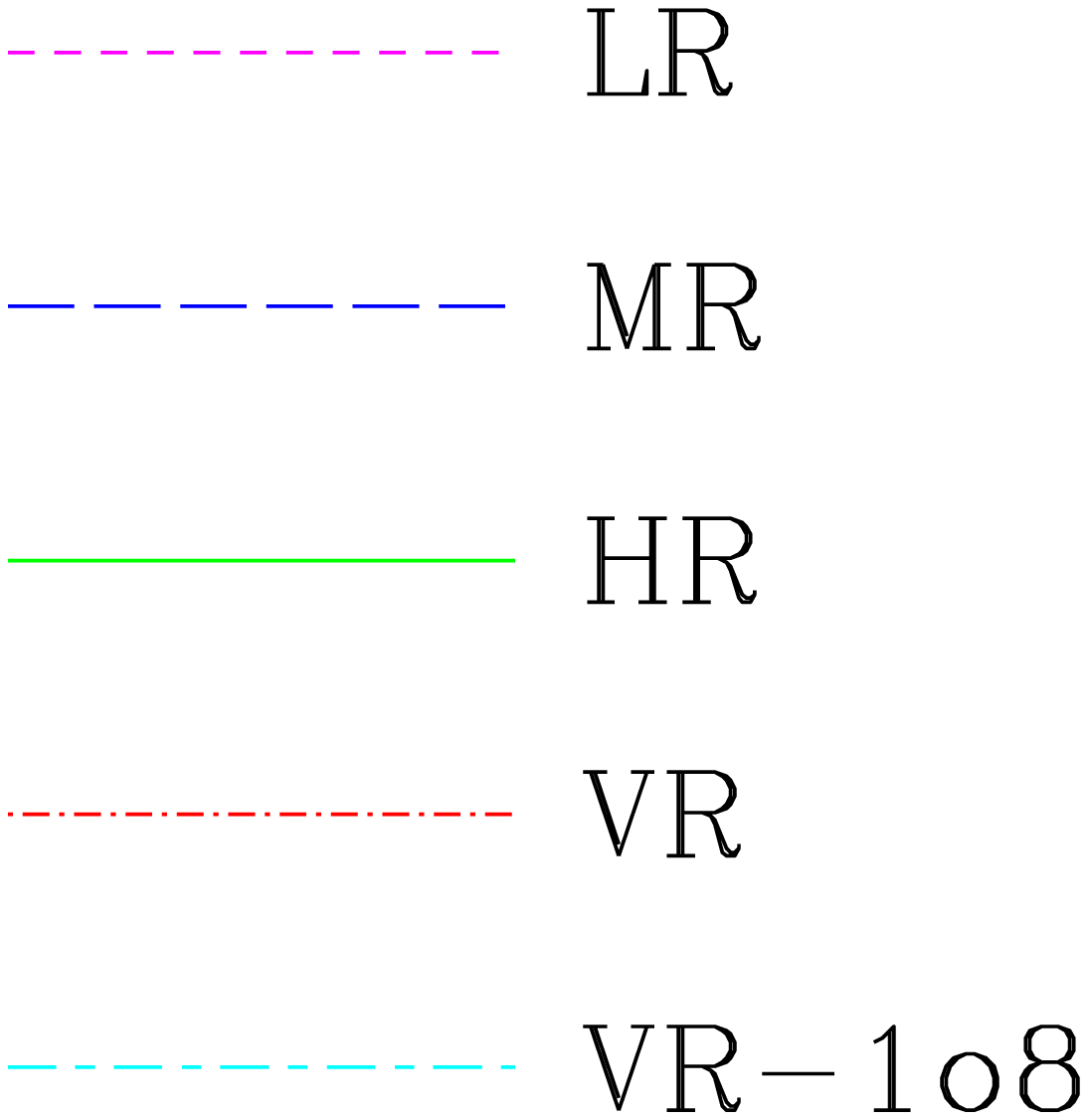,width=7.5cm} 
}
\hbox{
\psfig{file=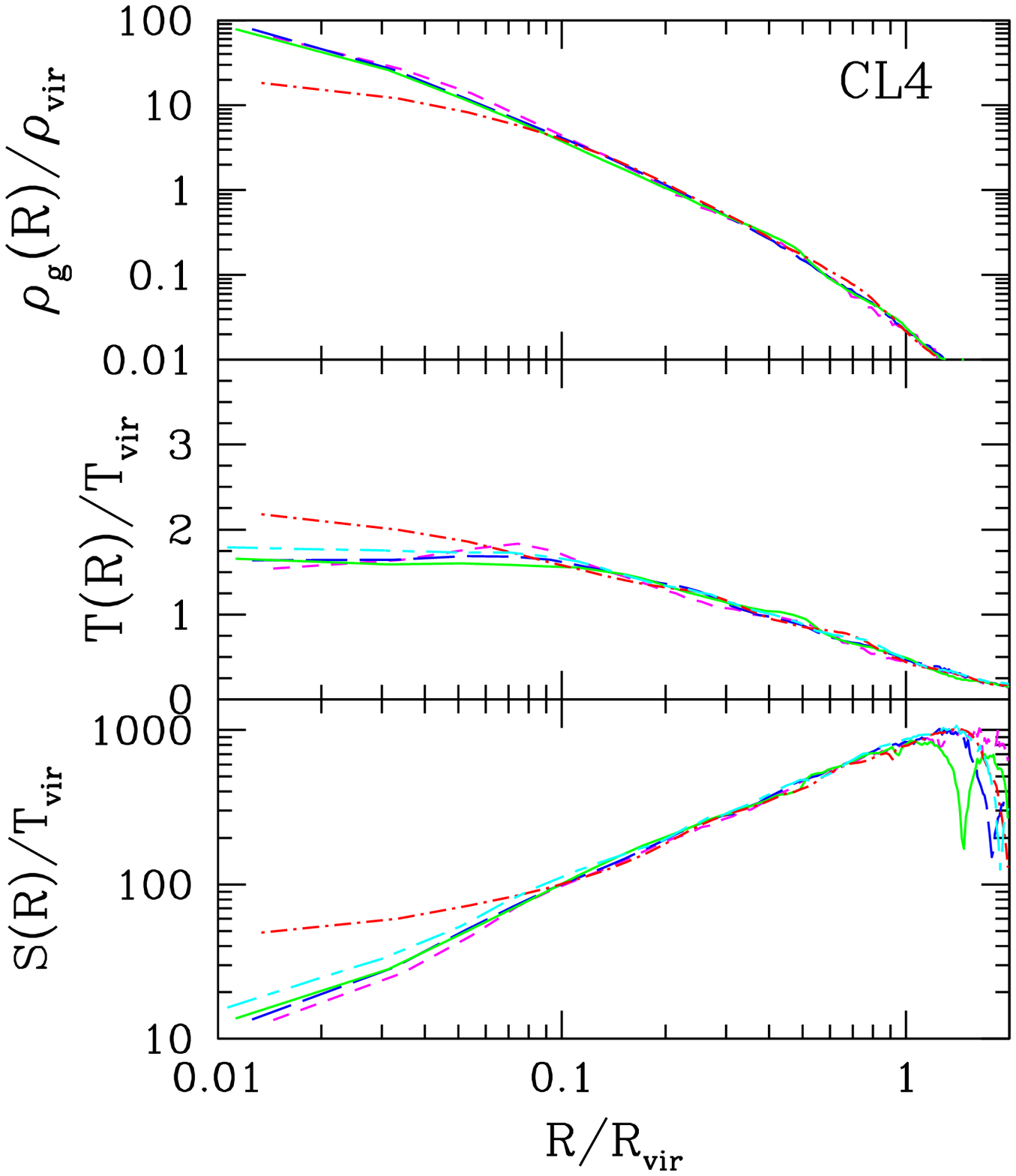,width=7.5cm} 
\psfig{file=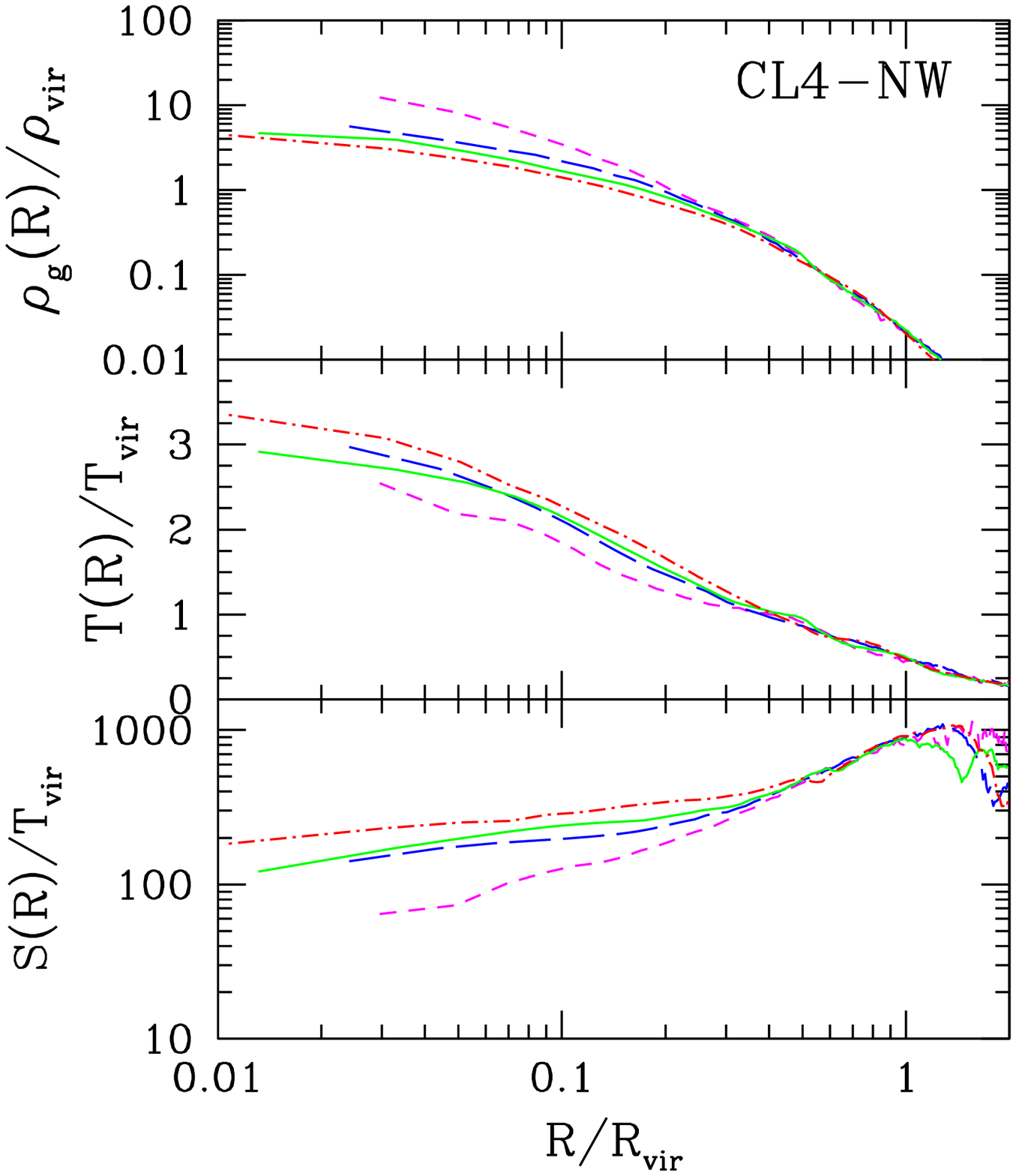,width=7.5cm} 
}
}}
\caption{The effect of resolution on the profiles of gas density,
  temperature and entropy (from top to bottom in each panel). Each panel shows
  the profiles for the different resolutions at which each cluster of 
  \tarkin\ has been simulated. For the CL4 cluster, we also show the results
  for the series of runs where the effect of winds has been excluded (NW).}
\label{fi:profs_res}
\end{figure*}

\begin{figure*}
\centerline{
\hbox{
\psfig{file=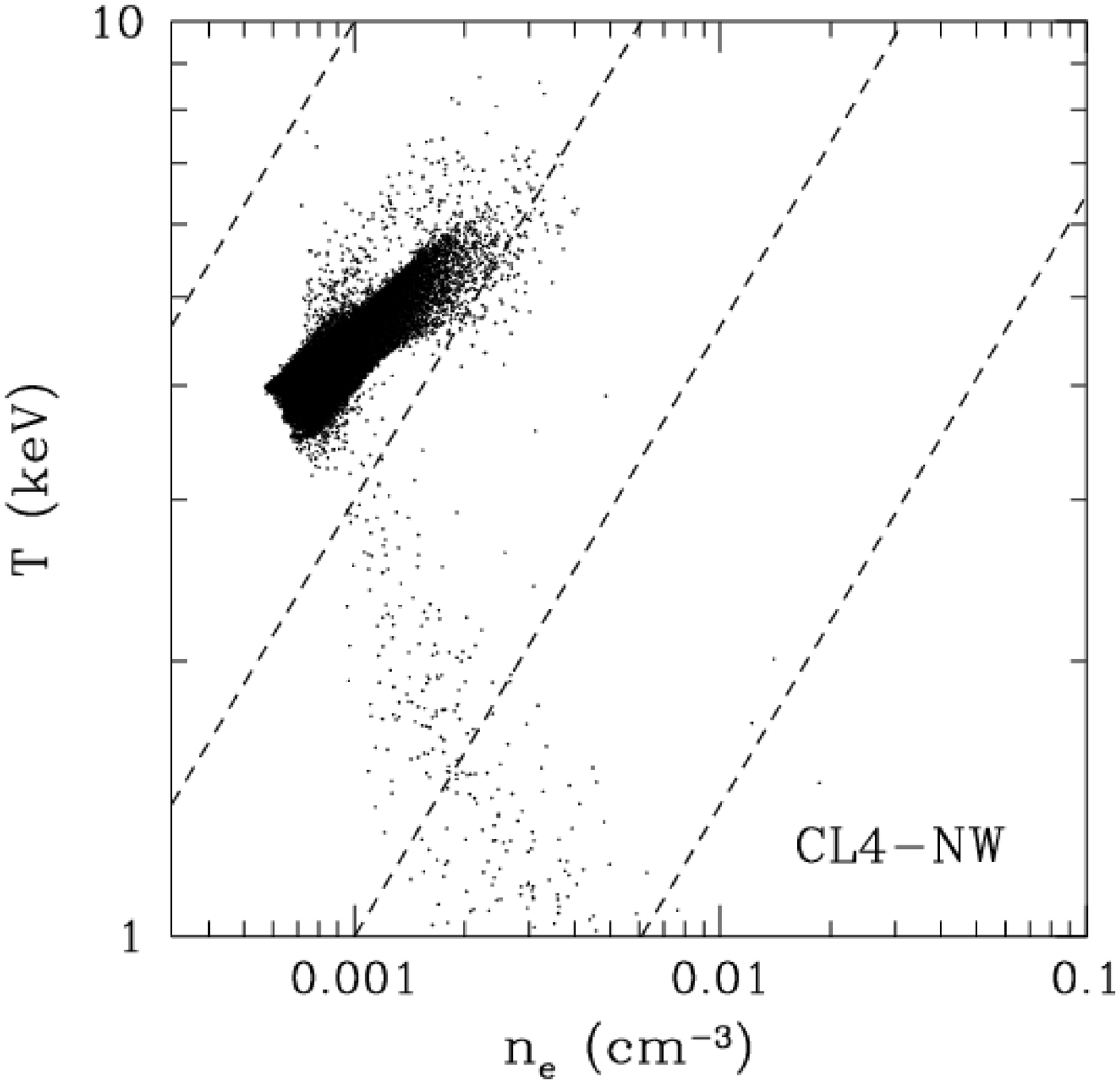,width=8.cm} 
\psfig{file=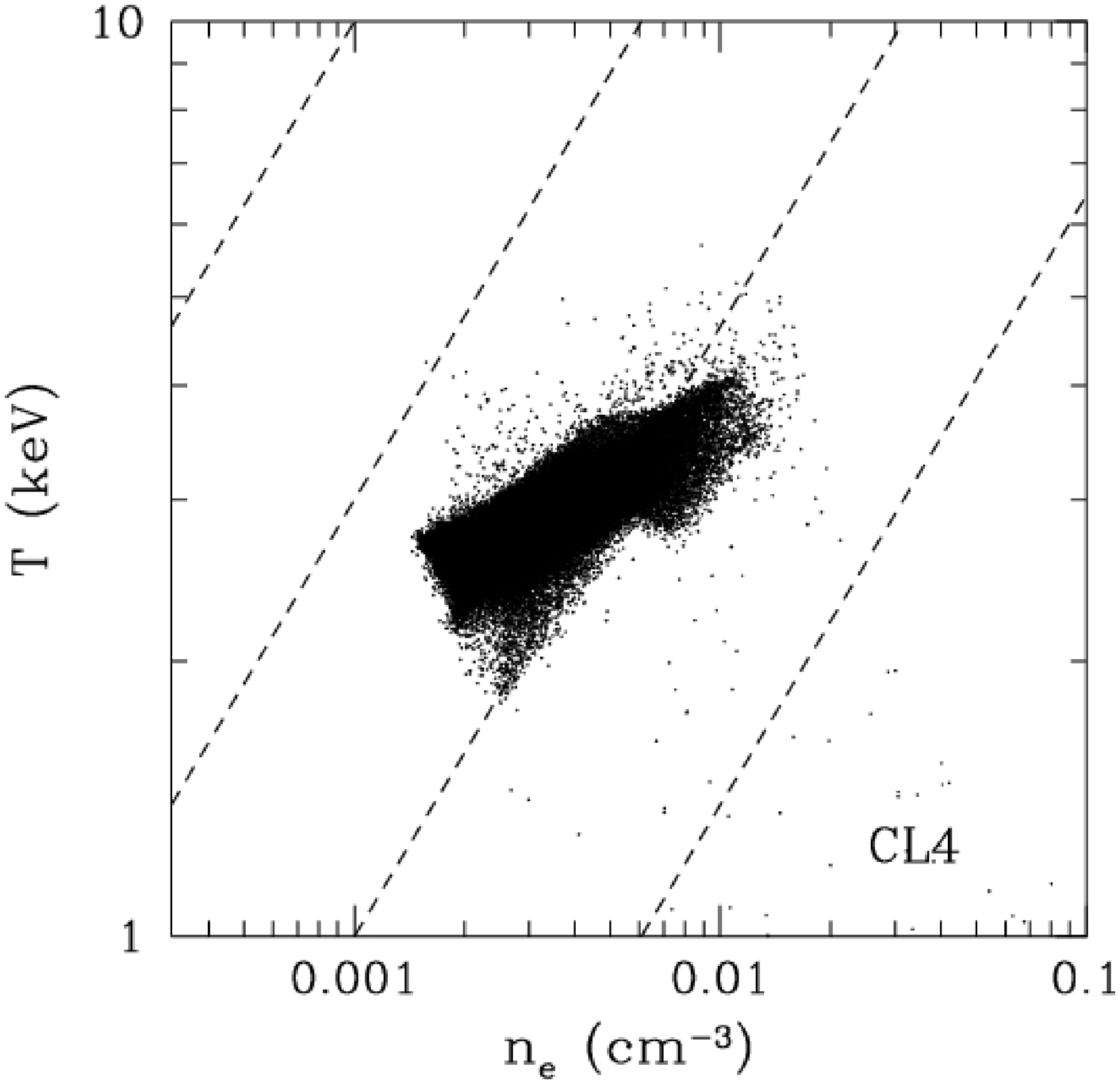,width=8.cm} 
}}
\caption{The distribution of gas particles for the very
  high--resolution (VR) version of the CL4 cluster, lying within $0.1\rvir$
  around the centre, in the $n_e$--$T$ plane ($n_e$ and $T$: electron number
  density and temperature of each gas particle, in units of cm$^{-3}$ and keV,
  respectively). Left and right panels are for the run without and with the
  effect of galactic winds, respectively. In each panel, the dashed curves
  indicate levels of constant entropy ($S=1000$, $300$, $100$, and $30$ keV
  cm$^2$, from upper to lower curves).}
\label{fi:phase}
\end{figure*}

How do our results compare with previous simulations?  An
overproduction of stars in simulations including radiative physics has
been a well established result for several years
\citep[e.g.,][]{1993ApJ...412..455K,1998ApJ...507...16S,2000ApJ...536..623L,2002ApJ...579...23D}.
This problem is generally thought to be solved by a physical feedback
process, which prevents excessive gas cooling with a continuous energy
supply. However, realistically modelling such a heating process is a
difficult problem. For instance, \cite{2003MNRAS.342.1025T} verified
that using thermal feedback instead of the kinetic one applied here,
the fraction of cooled baryons is of order 30 per cent, with no
evidence for convergence with better numerical resolution.
\cite{2002MNRAS.336..527M} claimed that overcooling within clusters
can be eventually avoided by suitably pre--heating gas at high
redshift.  \cite{2004MNRAS.355.1091K} implemented a scheme of thermal
feedback which increases the entropy of gas particles which are just
undergoing cooling. Although this feedback is tuned to reproduce
several X--ray observational properties of clusters, it still provides
a too large star fraction, of about $\simeq 25$ per cent. While all
these results are based on SPH codes, \cite{2005ApJ...625..588K} used
an adaptive Eulerian code, which also includes cooling, star formation
and thermal feedback. They find rather large stellar
fractions as well, with values of about 30--40 per cent within the
cluster virial radius, which is consistent with the SPH simulations.

\subsection{Thermodynamics of the hot gas}
In Figure \ref{fi:profs_res}, we show the radial profiles of gas
density, temperature, and entropy, for the the four clusters the
\tarkin, at different resolutions. We find that the profiles are
rather stable against resolution on scales $R\magcir 0.1R_{\rm
vir}$. On smaller scales, where the complex physics of cooling, star
formation and winds' feedback plays a significant role, we detect
significant systematic changes of the profiles with increasing
resolution. In general, the gas density and entropy profiles become
slightly shallower, and the temperature profiles somewhat steeper, as
the resolution is increased. The size of this effect changes from
object to object, depending on the different dynamical histories,
being more apparent for the CL1 cluster and rather small for the CL3
cluster.

Shallower gas--density and entropy profiles can be due either to more
efficient non--gravitational heating from energy feedback or to more
efficient radiative cooling. In the first case, the stronger heating
places gas on a higher adiabat, thereby preventing it from reaching
high density in the central halo region \citep[e.g.][and references
therein]{2001ApJ...546...63T}.  In the second case, a more efficient
cooling turns into a more efficient selective removal of low entropy
gas from the hot phase \citep[e.g.,][]{2001Natur.414..425V}. As a
result, the gas density decreases, while leaving behind the hot phase
at its relatively higher entropy. If cooling is the process that
governs the resolution--dependence of the profiles, we would expect a
more efficient star formation and a larger fraction of cooled gas at
increasing resolution. However, this is not found for the runs
including galactic winds (see Figs.~\ref{fi:fstar_res} and
\ref{fi:sfr_profs_res}). On the other hand, increasing the resolution
has the effect of increasing star formation efficiency and, therefore,
the amount of feedback energy released to the gas at high
redshift. Indeed, galactic winds are expected to be more efficient at
high redshift, when they can more easily escape from shallower
potential wells. Better resolution hence increases the gas heating
associated with high--redshift star formation.

The inner slope of the temperature profiles obtained in radiative
simulations is known to be significantly steeper than observed for
real clusters \citep[e.g.,][ cf. also
\citealt{2004MNRAS.355.1091K}]{2004MNRAS.348.1078B}.  This effect has
been interpreted as being due to the adiabatic compression of gas
which falls in from outer cluster regions, as a consequence of the
reduced pressure support after removal of cooled gas. Our results show
that improving resolution does not help to reconcile simulated and
observed temperature profiles. In fact, this discrepancy can become
even worse with better numerical resolution.

\begin{figure*}
\centerline{
\hbox{
\psfig{file=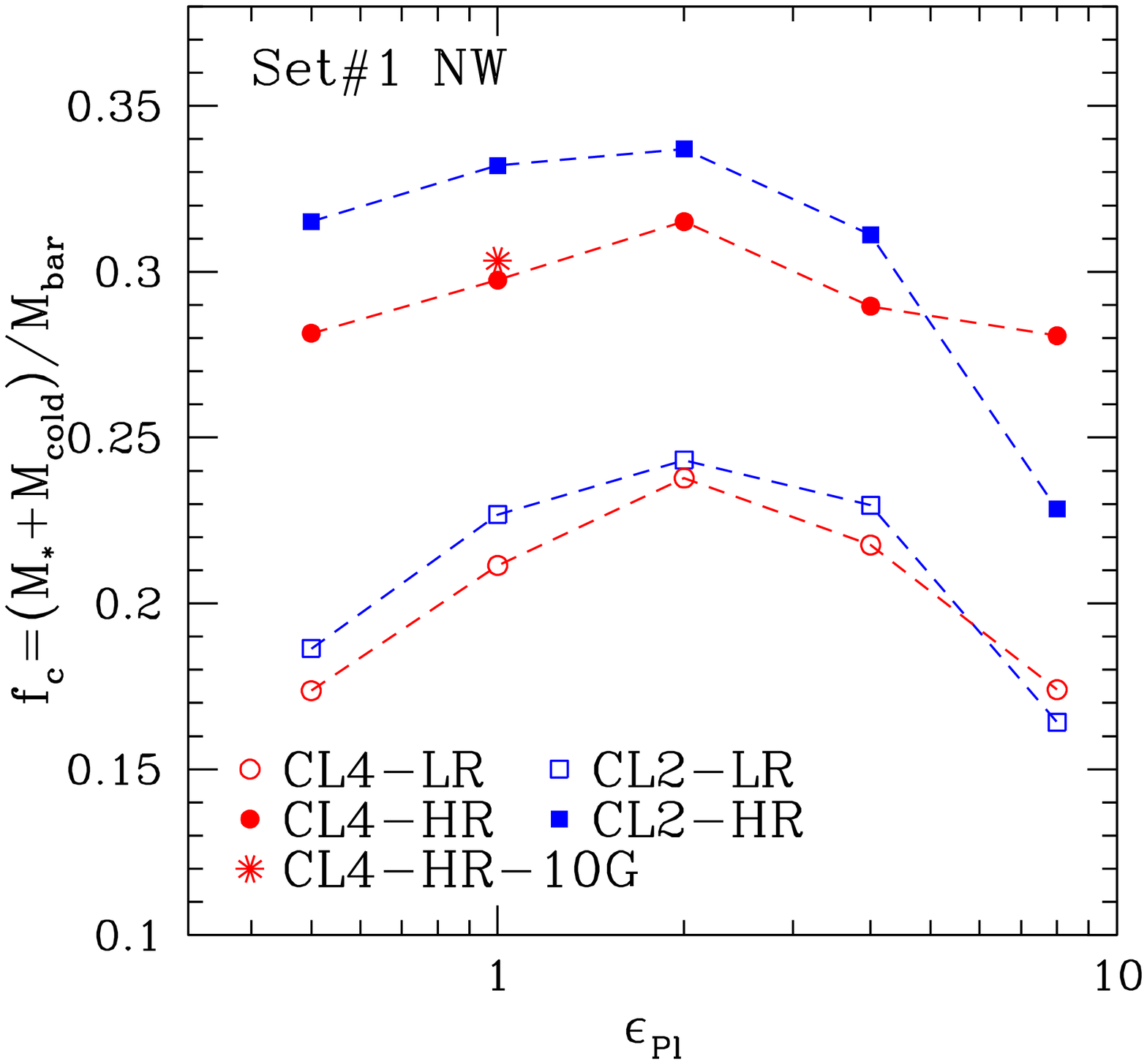,width=8.cm} 
\psfig{file=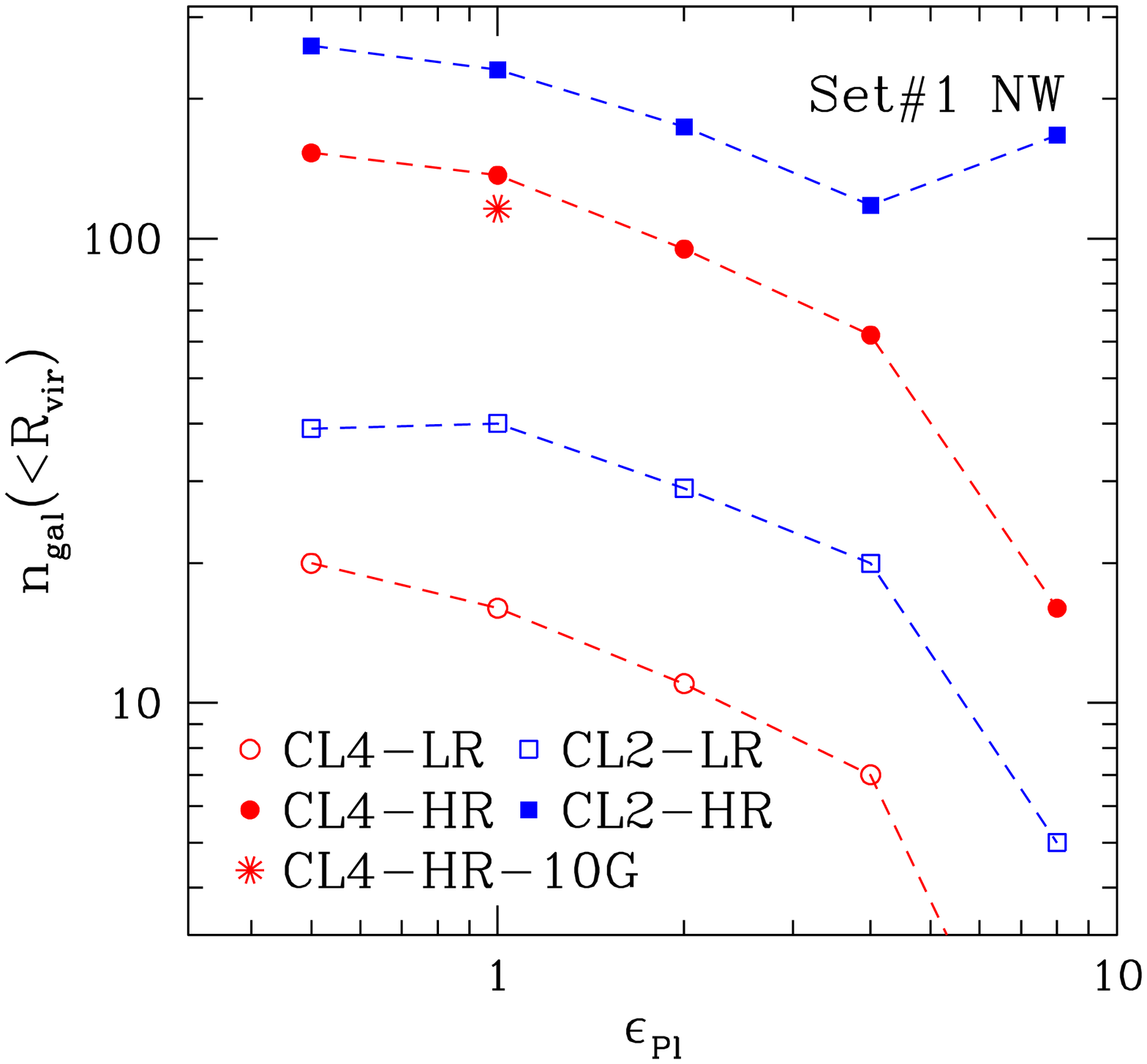,width=8.cm} 
}
}
\caption{The fraction of cooled baryons (left panel) and the number of
  galaxies (right panel) within the virial radius, as a function of the
  gravitational softening, for the CL2 and CL4 clusters (both in the LR and
  HR version). The softening is given in units of the value assumed for the
  reference runs (see Table~\ref{tab:res}). These simulations did not include
  galactic winds (NW runs).
}
\label{fig:fstar_soft}
\end{figure*}

In Fig.~\ref{fi:profs_res}, we show profiles of the gas properties for
the CL4 runs with no winds (NW). Also for these runs, we note similar
trends with resolution as for the runs including feedback by
winds. However, in this case the increase of the central entropy
level, and the corresponding decrease of the gas density, are
explained by the higher efficiency of cooling with better
resolution. Therefore, although the runs with and without feedback by
winds show similar trends with resolution, the interpretation of these
effects is completely different in the two cases.

Furthermore, a comparison of the CL4 runs with and without winds shows
that, at fixed resolution, the former have higher central density,
shallower temperature profiles and lower entropy levels.  In order to
explicitly demonstrate the effect of feedback on the gas
thermodynamics in the central cluster regions, we show in Figure
\ref{fi:phase} a phase diagram of the gas particles lying within the
central $0.1\rvir$ for the CL4 cluster, with and without feedback by
galactic winds. Perhaps somewhat counterintuitively, the effect of
feedback is that of lowering the typical temperature of the gas in the
central cluster regions, and shifting it to a lower adiabat. This
result emphasizes the role of feedback in the central cluster
regions: a continuous supply of energy by feedback has the effect of
keeping a population of relatively low--entropy gas particles in the
hot phase, which have a short enough cooling time that they would
otherwise have dropped out of the hot phase and cooled onto the
centre. The presence of such gas particles explains the higher gas
density and the lower entropy level. Furthermore, the higher pressure
support associated with the feedback energy reduces the compressional
heating of infalling gas and, as a consequence, the ICM temperature in
the central cluster regions.

Observational data on the temperature structure in cool--core clusters
indicate that the ICM reaches there a lower limiting temperature which is
about 1/2--1/4 of the overall cluster virial temperature
\citep[e.g.,][]{2001A&A...365L.104P,2001ApJ...560..194M,2002A&A...382..804B}.
The generally accepted interpretation is that some sort of feedback prevents
gas from reaching lower temperature by compensating its radiative losses, thus
suppressing also the cooling rate. Our results on the effect of feedback in
simulations clearly corroborates this picture. However, our implemented
feedback is still not efficient enough to suppress the cooling rate and,
correspondingly, the compressional heating to the observed level. As a
consequence, the temperature profiles in the central regions are still steeper
than observed, and no cool cores are created.

Finally, we consider the effect of changing $\sigma_8$ on the profiles of
gas--related quantities (right panel of Fig. \ref{fi:sig8}). We find that the
gas properties are left almost unchanged, once they are rescaled to compensate
for the difference in virial temperature. However, slightly shallower density
and entropy profiles, and a slightly steeper temperature profile characterize
the runs with $\sigma_8=0.9$. This can be understood as a consequence of the
enhanced cooling, which removes a larger fraction of gas from the hot phase in
the central cluster regions and, correspondingly, increases the compressional
heating of the in--flowing gas.

\section{The effect of numerical heating}

\subsection{Changing the gravitational softening}
The choice for the gravitational softening is a compromise between the
goal of resolving the smallest scales possible and the need to ensure
that the relaxation time by two--body interactions is much larger than
the typical age of the simulated structure
\citep[e.g.][]{1992MNRAS.257...11T}.  Using a very small softening may
allow one to resolve a larger number of small halos, but this comes at
the price of introducing spurious numerical heating of the gas, which
could artificially suppress cooling. Therefore, by increasing the
softening we expect to first see a more efficient cooling, until a
point is reached when only large halos, with their comparatively long
cooling times, are resolved and the cooling efficiency declines
again. This implies that a softening value should exist which
maximizes the amount of cooled gas, while smaller and larger values
lead to less cold gas as a result of spurious gas heating or lack of
resolution, respectively.

To check for this effect, we show in Figure~\ref{fig:fstar_soft} the
dependence of the fraction of cooled baryons, $f_c$, and of the number
of identified galaxies $n_{\rm gal}$ within the virial radius on the
adopted gravitatiobal softening length. We show results both for the
low--resolution (LR) and the high--resolution (HR) versions of the CL2
and CL4 clusters. In order to avoid mixing the effects of numerical
heating with that of efficient feedback, we have performed these
simulations by switching off feedback by galactic winds (NW). In the
plot of Figure~\ref{fig:fstar_soft}, the softening is given in units
of that adopted for our reference simulations, as reported in Table
\ref{tab:res}. The results clearly confirm our expectation: both a too
large and a too small softening lead to a decrease of the amount of
cooled gas. Quite interestingly, the softening (in units of the
reference value) at which the cooled baryon fraction is maximised is
the same at low and high resolution.  This demonstrates that the
adopted scaling of the softening length with mass resolution,
$\epsilon_{\rm Pl}\propto m_{\rm gas}^{1/3}$ is in fact a reasonable
choice.

In general, we find that the maximum value of $f_c$ is attained
  for a softening value about twice as large as our reference
  value. This implies that a two times larger softening should then be
  preferred, although the difference in $f_c$ when using
  $\epsilon_{\rm Pl}=1$ instead of 2 is always rather small, especially
  at high mass resolution.

  As for the number of identified galaxies, it monotonically decreases
  with increasing softening (right panel of
  Fig. \ref{fig:fstar_soft}). Using $\epsilon_{\rm Pl}=2$ provides a
  $\sim 30$ per cent smaller number of galaxies than for
  $\epsilon_{\rm Pl}=1$. However, this argument does not necessarely
  represent a valid support for the choice of a smaller
  softening. Indeed, a spurious energy transfer from DM to gas may
  lead to a tightening of the DM halos, thereby producing a larger
  number of small halos. Clearly, a close investigation of the optimal
  strategy to minimize the effect of numerical heating would require a
  more detailed analysis than that presented here, including a close
  investigation of the mass function of the resolved galaxies. We plan
  to present this analysis in a forthcoming paper.

We finally note
that the anomalous number of galaxies for the HR version of the CL2
cluster is apparently a result of a misclassification by SKID, as we
checked by visual inspection.  Here, the distribution of star
particles shows that they are arranged in a few over-merged galaxies,
most of which have pretty large tidal tails. The SKID algorithm
sometimes splits these tails into separate gravitationally bound
structures, which are then misclassified as galaxies.

\begin{figure}
\centerline{
\includegraphics[width=0.45\textwidth]{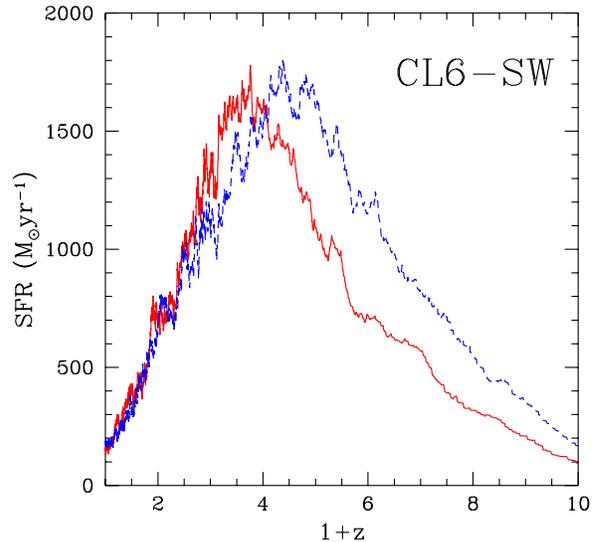}
}
\vspace{-2.5truecm}
\caption{Comparison of the results for the CL6 cluster using both
  $z=5$ (solid line) and $z=2$ (dashed line) for the transition
  redshift from physical to comoving softening.  Strong winds (SW)
  have been included in these simulations.}
  \label{fi:soft_tr}
\end{figure}

As a further test, we have compared the results of the CL6 run
  with strong winds, assuming $z=2$ and $z=5$ for the redshift of
  transition from comoving to physical softening. As shown in Figure
  \ref{fi:soft_tr} the smaller high--$z$ softening provides an
  enhanced efficiency of star formation. In this case, the cooling
  efficiency within small resolved halos dominates over the effect of
  numerical heating in defining the pattern of star formation. 

\subsection{Gas--DM particle splitting}

%It is standard practice in cosmological SPH simulations to add gas particles
%to DM-only initial conditions by splitting each parent particle into a DM one
%and a gas one, whose mass ratio is chosen to match the cosmic baryon fraction.
%This implies that the simulation needs  to follow the gravitational interaction
%of two particle populations of rather different particle masses. As discussed
%by \cite{1997MNRAS.288..545S}, this causes a spurious heating of the gaseous
%component as a consequence of the two--body relaxation process between the
%heavier dark matter and the lighter gas particles. This numerical
%heating induces an artificial suppression of the cooling efficiency within
%small collapsing halos. 

As we have already discussed, heating of gas particles from
two--body encounters may spuriously alter the thermodynamic properties
of the gas. Based on an estimate of the heating time from two--body
encounters in the impulse approximation \citep{1987gady.book.....B},
\cite{1997MNRAS.288..545S} analytically derived the heating time
associated with such encounters. Under the assumption of hydrostatic
equilibrium, this heating time is inversely proportional to the mass
of the DM particle, while being independent of the gas particle
mass. As a consequence, a limiting mass of DM particles can be derived
for two--body heating rate to dominate over the cooling rate within a
given halo, as a function of the virial temperature of the halo
itself.  According to their results, the mass resolution achieved in
our runs, with the possible exception of the low--resolution (LR)
series for clusters of \tarkin, should be adequate to prevent spurious
heating within halos having temperature of at least few keV.  While
this holds for already formed halos at low redshift, the situation may
be different at high redshift, when gas cooling takes place within
small, just resolved galaxy--sized halos.  Also, in the first
generation of halos, substantial residual gas motions exist which
could give rise to an additional artificial energy transfer in
two-body encounters from the heavier DM particles to the lighter gas
particles.

In order to reduce the amount of numerical heating from such energy
equipartition effects one may try to increase the number of DM
particles, thereby decreasing their mass, while keeping the mass
resolution of the gas fixed. In order to check for this effect, we
resimulated the CL4 halo assuming very high resolution (VR) for the DM
particles. However, in this case we generated initial displacements
for an eight times smaller number of gas particles (1o8 runs in Table
\ref{tab:tests}). In this way, gas and DM particles have a comparable
mass: $m_{\rm gas}/m_{DM}\simeq 1.06$ instead of $\simeq 0.15$. In
terms of gas--mass resolution, this simulation is intermediate between
the MR and HR runs, but with a more accurate treatment of gravity,
thanks to the larger number of DM particles, which should suppress
numerical heating.

In Fig.~\ref{fi:fstar_res}, we show the effect on the cooled gas
fraction and on the number of formed galaxies with the
asterisks. Quite interestingly, the resulting cooled fraction is very
similar to that obtained for the MR and HR runs of the same
cluster. 
%This indicates that the total amount of cooled gas at $z=0$
%in these runs is regulated by the mass of the gas particles rather
%than by that of the DM particles. 
As for the number of resolved
galaxies, it is close to that obtained for the corresponding HR run,
which has anyway a better gas--mass resolution. This increase in the
number of identified galaxies inside the cluster can have two possible
explanations. It is either due to an intrinsic increase of the number
of resolved halos where gas cooling takes place, or to an improved
capability of small halos to survive disruption in the cluster tidal
field. To check this, we compare the values of $n_{\rm gal}(<R_{\rm
vir})$ inside the cluster with the corresponding number of galaxies in
the region $R_{\rm vir}<R<4R_{\rm vir}$ around the cluster. Within
\rvir, we identify 56 galaxies for the HR run and 53 galaxies for the
1o8 run.  However, the difference between the two runs increases
significantly in the outer cluster regions, with 123 galaxies found
for the HR run and 75 for the 1o8 run.  This shows that the better
gas--mass resolution of the HR run in fact produces a larger number of
galaxies in the field, while the improved accuracy in the
gravitational force computation for the 1o8 run compensates for this
effect within clusters, where it plays a significant role in
preventing tidal disruption of small galaxies.

With respect to the cosmic star--formation history, the 1o8 run (heavy
solid curve in Fig.~\ref{fi:sfr_profs_res}) shows a behaviour which is
not strictly intermediate between MR and HR runs. The onset of star
formation takes place at a higher redshift than in the HR runs, and
then proceeds in a more efficient way. Quite remarkably, the peak of
star formation is even higher than for the highest resolution (VR)
run.  Finally, results of the 1o8 run for the gas profiles are shown
with the long-short--dashed curve in the bottom left panel of
Fig.~\ref{fi:profs_res}. Much like for the results on the cooled gas
fraction, the resulting profiles are very similar to those of the MR
and HR runs.

Overall, the test on the degree of spurious gas heating from two--body
relaxation confirms that a DM particle mass of at most $\simeq
10^9\msun$ is sufficient to provide a reliable description of gas
cooling within an already formed cluster of galaxies at low
redshift. However, increasing the DM mass--resolution in order to 
reduce numerical heating of the gas from two--body encounters has a
non--negligible effect on the number of resolved galaxies and on the
star formation history, as a result of the better resolution of early
structure formation.  The comparatively low computational cost of the
gravity part in a hydrodynamical simulation with radiative cooling,
and the availability of large amounts of memory in modern
supercomputers, may make it attractive to adopt a larger number of DM
particles than gas particles in future simulation work.

\begin{figure*}
\centerline{
\hbox{
\psfig{file=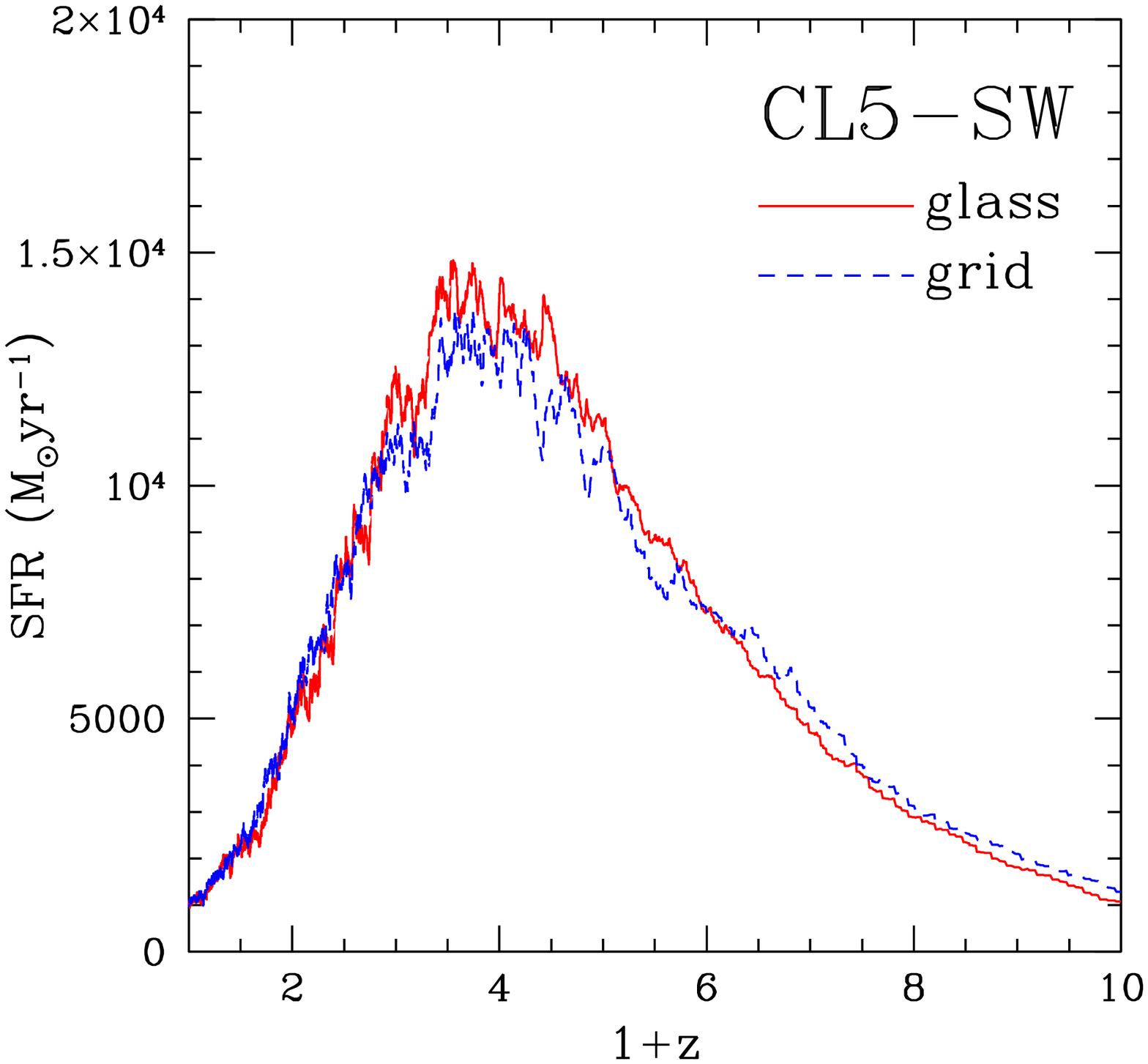,width=8.cm} 
\psfig{file=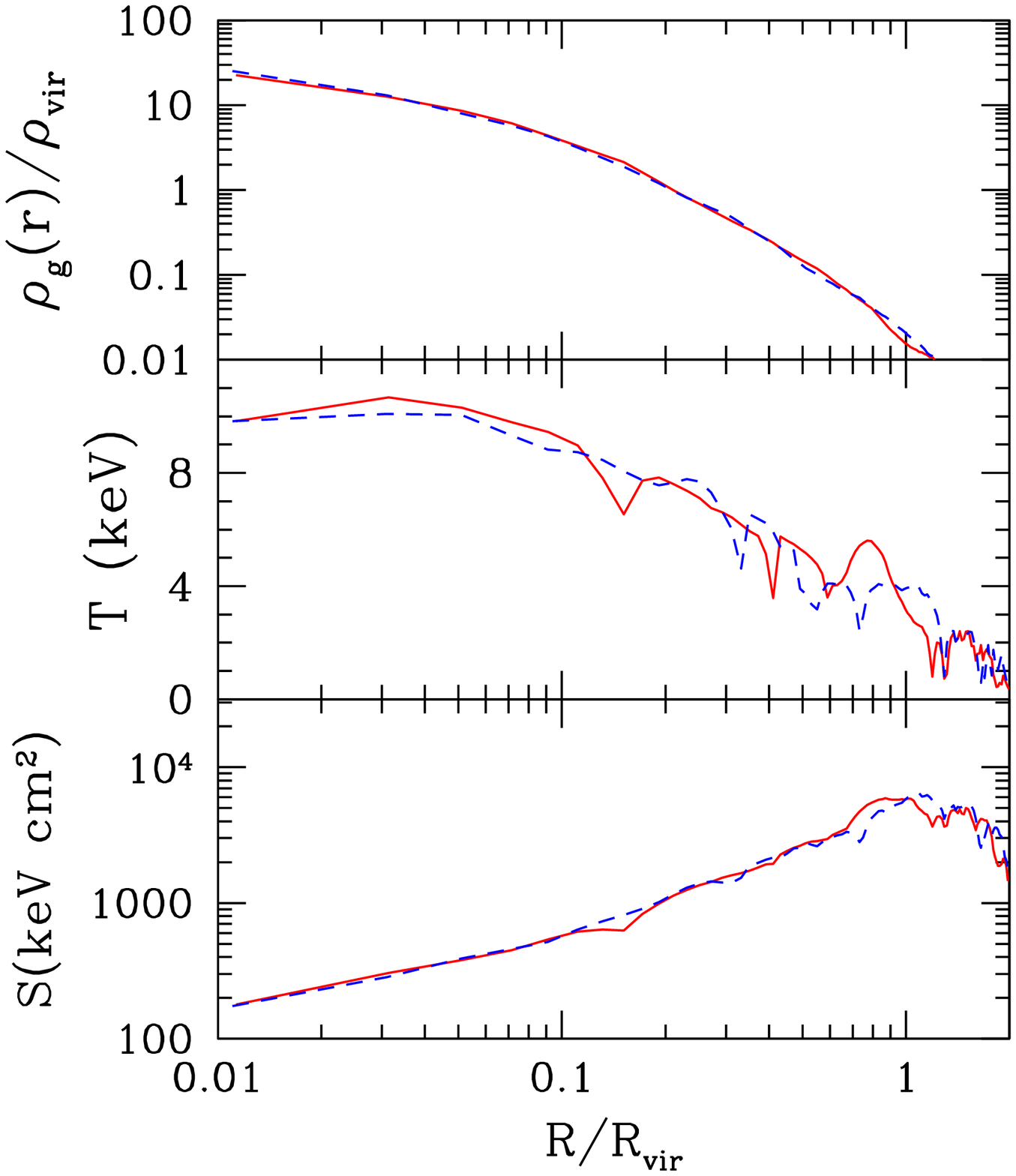,width=8.cm} 
}}
\caption{Comparison of the results for the CL5 cluster 
  using both glass (solid lines) and grid (dashed lines) initial conditions.
  The left panel is for the redshift dependence of the star formation rate,
  while the right panels show the profiles of gas density, temperature, and
  entropy. Strong winds (SW) have been included in these simulations.  }
  \label{fi:grid_vs_glass}
\end{figure*}

\subsection{Grid vs. glass initial conditions}

As we discussed in Section~\ref{sec:sims}, the initial Lagrangian particle
distributions of the simulations of \tarkin\ have been realized as a grid,
while those of \hutt\ have been a glass.  The glass-based technique to
generate initial conditions (ICs; \citealt{WH96.1b}) aims at suppressing
effects due to the regularity of the initial grid, which amplifies
structure at the scale of the mean-interparticle separation.  In the glass
scheme, particle positions are initially generated randomly in the simulation
box, but are then evolved backwards in time until they reach an amorphous,
minimum energy configuration where each particle experiences only vanishingly
small forces. The resulting irregular particle distribution lacks prefered
directions and should be less affected by the symmetries that occur in
the grid method.

In order to check the effect of using either one of the two techniques
to generate ICs, we have rerun the CL5 cluster with strong winds (SW),
but this time starting from grid displacements. The results of this
test are shown in Figure~\ref{fi:grid_vs_glass}, where we plot the
corresponding star--formation histories (left panel) and the radial
profiles of gas properties (right panels). This comparison
demonstrates that, at least at the resolution relevant for our
simulations, the difference between using grid or glass ICs is very
small.  Looking at the details of the comparison, it turns out that
the grid--based run has a slightly higher star--formation at $z\magcir
6$.  This may be due to a contribution of the small--scale fluctuation
modes around the Nyquist frequency, which should collapse more
efficiently in the grid case (and part of this may be artificial),
favoring somewhat earlier cooling at high--$z$. However, the star
fractions within the cluster virial radius at $z=0$ are $f_*=0.17$ for
both runs. The number of identified galaxies within the same region is
412 and 380 for the grid and glass runs, respectively. This appears to
confirm that grid--based ICs show slightly higher power on small
scales, which, in turn, generates a slightly larger number of
galaxies. As for the profiles of gas related quantities, we note that
they also overlap quite closely. The only noteworthy difference is
that the positions of merging substructures vary. However, such
differences in orbital timing are expected and common when different
methods for the generation of ICs are used.

The general result of this comparison is that, at least at the
resolution relevant for our cluster simulations, the effect of using
either grid or glass ICs is very small, in any case negligible with
respect to other numerical effects that we have explored in this
paper.

\section{Conclusions} \label{sec:conc}

In this paper we have presented results from a large set of
hydrodynamical simulations of galaxy clusters, carried out with the
Tree+SPH code {\small GADGET-2}. Our simulations include radiative
cooling, star formation and energy feedback by a phenomenological
model for galactic winds.  The main target of our analysis has been
the study of the stability of simulation results with respect to
numerical parameters, such as mass resolution or gravitatiobal
softening length. We also considered different sources of numerical
heating, and their interplay with the complex physical effects
included. As such, our analysis also represents a validation study of
our previous results \citep[e.g.,][]{2004MNRAS.348.1078B}, which were
based on a large cosmological box simulated a relatively low
resolution.

Our simulated clusters span more than one order of magnitude in
collapsed mass and several decades in mass resolution. At the highest
resolution, the mass of the gas particles is $m_{\rm gas}\simeq
1.5\times 10^7 h^{-1}{\rm M}_\odot$, which allows us to resolve the
virial region of a Virgo--like cluster with more than 2 million gas
particles and at least as many dark--matter (DM) particles. Our main
results are concerned with the effects of resolution on the properties
of the stellar populations and on the intra--cluster medium of the
simulated galaxy clusters. They can be summarized as follows.

\begin{description}
\item[(a)] In the absence of an efficient energy feedback, the
  fraction of cooled baryons steadely increases with resolution,
  reaching $\simeq 35$ per cent at the highest achieved resolution (VR
  runs in Table \ref{tab:res}), with no indication for convergence.
\item[(b)] Including feedback from galactic winds has the effect of
  stabilizing the stellar fraction inside clusters. Assuming a high
  efficiency for the supernova driven winds of order unity, we find
  that the fraction of cooled baryons converges already at a
  relatively modest resolution, with an indication to decrease very
  slightly at the highest resolution. This arises as a consequence of
  the self--regulation property of star formation and feedback. While
  improving the resolution increases the star formation efficiency at
  very high redshift, this at the same time also provides a
  significant contribution to gas pre--heating which reduces star
  formation later on.  The fraction of cooled baryons within \rvir~
  lies in the range 12-18 per cent, with a decreasing trend with
  cluster mass. These values increase by about 15 per cent when the
  normalization of the power--spectrum is raised from $\sigma_8=0.8$
  to 0.9.
\item[(c)] The feedback provides the necessary continued energy supply
  to keep gas particles of comparately low entropy and short cooling
  times in the hot phase of clusters, while without feedback these
  particles would cool down and drop out of the cluster atmosphere. As
  a result, the central gas density is higher in runs with feedback
  than in the runs with no galactic winds. The temperature profiles
  are shallower in the central cluster regions, while isentropic cores
  are much less pronounced, thus alleviating the discrepancy with the
  observed properties of the intra--cluster medium.
\end{description}

A further series of tests presented in this paper concerns the effect
of numerical heating. The main results from these tests can be
summarized as follows.

\begin{description}
\item[(a)] Our Plummer--equivalent force softening of $\epsilon_{\rm
  Pl}\simeq 10\hk$ at a mass resolution of $m_{\rm DM}\simeq 5\times
  10^9\msun$, scaled to other particle masses as $\epsilon_{\rm
  Pl}\propto m_{\rm DM}^{1/3}$, represents a reasonable compromise
  between the need of preventing a spurious numerical heating of the
  gas, and the desire to resolve galaxies in the largest possible
  number of small DM halos.
\item[(b)] 
%Reducing the mass of the DM particles such that $m_{\rm gas}\simeq
%m_{\rm DM}$ 
{Increasing the mass resolution in the DM component while keeping the gas
mass resolution fixed,} helps to reduce numerical gas heating by
two--body encounters \citep{1997MNRAS.288..545S}. We find that
decreasing $m_{\rm DM}$ has a negligible effect on the fraction of
cooled gas at $z=0$ and on the radial profiles of gas
properties. However, it has a non--negligible effect on the star
formation history, which occurs with enhanced efficiency at $z\magcir
3$. This is the result both of a non--negligible numerical heating in
small halos when gas and DM particles have rather different masses,
and also of a better resolved dark matter content in the first
generation of halos.
\end{description}

In sum, our analysis helps to establish and delineate the regime of
numerical reliability of the present generation of hydrodynamical SPH
simulations of galaxy clusters. Even though quite technical in nature,
it is clear that systematic tests like the ones discussed here are
required to elucidate the often delicate interplay betweeen numerical
effects and the physical model that is studied. Only when these
effects are understood, the predictive power of numerical experiments
can be fully exploited.  At the same time, this understanding is also
required to successfully push to new generations of simulations with
yet higher resolution. A reassuring aspect of the results discussed
here is that even simulation models with complex and highly non-linear
physical models for star formation and feedback can produce
surprisingly robust results. This clearly is an encouragement for
attempts to improve the fidelity with which the physics is represented
in future simulation work.

\section*{acknowledgements}
The simulations were carried out on the IBM-SP4 machine and on the IBM-Linux
cluster at the ``Centro Interuniversitario del Nord-Est per il Calcolo
Elettronico'' (CINECA, Bologna), with CPU time assigned under INAF/CINECA and
University-of-Trieste/CINECA grants, on the IBM-SP3 at the Italian Centre of
Excellence ``Science and Applications of Advanced Computational Paradigms'',
Padova and on the IBM-SP4 machine at the ``Rechenzentrum der
Max-Planck-Gesellschaft'', Garching. This work has been partially supported by
the PD-51 INFN grant. We wish to thank an anonymous referee for
his/her comments which helped improving the presentation of the results.

\bibliographystyle{mn2e}
\bibliography{master}

\end{document}